\documentclass[showpacs,twocolumn,prb]{revtex4-1}

\usepackage{amsmath}
\usepackage{amssymb}
\usepackage{graphicx}
\usepackage{amssymb}
\usepackage{graphics}
\usepackage{epsfig}
\usepackage{color}
\usepackage{soul}
\usepackage{natbib}
\usepackage{bibentry}
\usepackage{appendix}

\newcommand{\ybfep}{YbFe$_5$P$_3$}

\begin{document}

\title{Quantum Critical Scaling in Quasi-One-Dimensional YbFe$_5$P$_3$}
\author{E. D. Bauer$^{1}$, K. E. Avers$^{1,2}$, T. Asaba$^1$, S. Seo$^1$, Y. Liu$^1$, A. Weiland$^1$, M. A. Continentino$^3$, J. M. Lawrence$^1$, S. M. Thomas$^1$, P. F. S. Rosa$^1$, J. D. Thompson$^1$,  and F. Ronning$^4$ }
\affiliation{$^{1}$MPA-Q, Los Alamos National Laboratory, Los Alamos, New Mexico 87545, USA\\
$^{2}$Department of Physics and Astronomy, Northwestern University, IL, 60208 USA\\
$^{3}$Centro Brasileiro de Pesquisas Físicas, Rio de Janeiro, 22290 Brazil\\
$^{4}$Institute for Materials Science, Los Alamos National Laboratory, Los Alamos, New Mexico 87545, USA}
\date{\today}

\begin{abstract}
We report measurements of the low temperature magnetization $M$ and specific heat $C$ as a function of temperature and magnetic field of the quasi-one-dimensional spin chain, heavy fermion compound \ybfep, which resides close to a quantum critical point.  The results are compared to the predictions of scaling laws obtained from a generalized free energy function expected near an antiferromagnetic quantum critical point (AFQCP). The scaling behavior depends on the dimensionality $d$ of the fluctuations, the coherence length exponent $\nu$, and the dynamic exponent $z$. The free energy treats the magnetic field as a relevant renormalization group variable, which leads to a new exponent $\phi=\nu z_h$, where $z_h$ is a dynamic exponent expected in the presence of a magnetic field. When $z_h=z$, $T/H$ scaling is expected, as observed in several compounds close to a QCP; whereas in \ybfep, a $T/H^{3/4}$ dependence of the scaling is observed. This dependence reflects the relationship $z_h=(4z/3)$ and a field exponent $\phi =4/3$. A feature of the scaling law is that it restricts the possible values of the exponents to two cases for \ybfep:  $d$=1, $\nu$=1, $z$=1, and $d$=2, $\nu$=1/2, $z$=2. 
\end{abstract}
\maketitle

\section{INTRODUCTION}

The investigation of quantum phase transitions and quantum critical phenomena is important in condensed matter.  A wide array of unusual behavior and interesting ground states are believed to arise from the strong quantum fluctuations near a quantum critical point (QCP),\cite{Sachdev2011} including: unconventional superconductivity in $f$-electron heavy fermion, high-T$_c$ cuprate and iron-pnictide compounds,\cite{Scalapino2012,Monthoux2007} strange-metal behavior in “twisted” dichalcogenide materials,\cite{Ghiotto2021} and quantum critical behavior in paraelectrics near a ferroelectric QCP.\cite{Rowley2014} Of particular interest is the study of quantum criticality in $f$-electron metals associated with a magnetic transition  that is tuned to zero temperature by the application pressure, magnetic field, or chemical substitution.\cite{Stockert2011,Stewart01,vonLohneysen07,Paschen2021} The quantum critical behavior of most magnetic heavy fermion materials is that of a mean-field quantum critical point (QCP).\cite{vonLohneysen07,Millis93} As initially pointed out by Hertz,\cite{Hertz76}  for $T=0$ quantum phase transitions, the dynamic exponent $z$, which governs the energy scale ($\omega$) of the fluctuations (i.e., $\omega \sim \xi^{-z}$, where $\xi$ is the correlation length), enters the free energy as additional degrees of freedom so that the effective dimensionality becomes $d_{eff} = d+z$. When this dimension becomes greater than the upper critical dimension $d_c=4$, the behavior of the system is Gaussian; for the marginal case with $d_{eff}$ =4, the behavior is Gaussian with logarithmic corrections.\cite{Millis93,vonLohneysen07}  The dynamic exponent is generally fixed by dynamical constraints, with $z$ typically being 2 for antiferromagnets and 3 for ferromagnetic metals. Because most of the heavy fermion compounds whose QCPs have been studied by neutron scattering are three-dimensional (3D), for $z=2$ or 3, we have  $d_{eff}> d_c$ yielding Gaussian behavior and breakdown of hyperscaling. Gaussian behavior is not only expected but is typically observed. For example, alloys of CeRu$_2$Si$_2$  and CeNi$_2$Ge$_2$  exhibit the $T^{3/2}$ behavior of the energy linewidth $\Gamma$ of the dynamic susceptibility\cite{Kadowaki2006,Wang2011,vonLohneysen07} $\chi{^{\prime\prime}}$ predicted by Hertz, Millis, and Moriya  (HMM) for 3D antiferromagnetic QCPs.\cite{Millis93,Hertz76,Moriya95,vonLohneysen07} Under these circumstances, the study of QCPs beyond the classical or Gaussian domain requires low-dimensional materials.

One quasi-1D material close to a QCP is \ybfep.\cite{Asaba2021} The orthorhombic structure of \ybfep{} is comprised of chains of Yb ions along the b-axis, as shown in the inset of Fig. \ref{props}a with an Yb-Yb distance of d$_{Yb-Yb}$ = 3.65 \AA, close to twice the metallic radius (2$R_{Yb} =2 \cdot 1.74$ \AA{} = 3.48 \AA), which is much smaller than the inter-chain distance of 5.6 \AA. This is expected to confer quasi-one-dimensional character to the magnetic fluctuations in this system, which is also reflected in the electronic band structure.\cite{Asaba2021} The large increase in specific heat and magnetic susceptibility below 10 K, where C/T saturates to a value $\sim$ 1.4 J mol$^{-1}$K$^{-2}$ below 0.5 K with no magnetic order observed above 0.08 K, indicates close proximity to a QCP. Above 0.5 K, the electrical resistivity is linear with temperature but at lower temperatures it approaches T$^2$ behavior. Under applied pressure (P), \ybfep{} shows no sign of magnetic order up to 2.5 GPa and above 0.25 K.  The resistivity follows a linear temperature dependence, $\rho$(T) = $\rho_0 + AT^{\eta}$ ($\eta$ = 1) with an increasing $A$ coefficient with $P$,  suggesting a QCP near 3 GPa.\cite{Asaba2021} In addition, when \ybfep{} is alloyed with Co or Ru, a QCP is observed at a critical concentration $y_c$ beyond which the compound becomes antiferromagnetic.\cite{Avers2024}  Given this behavior, \ybfep{} is clearly a candidate for low-dimensional, non-Gaussian behavior near a QCP.

One method for studying the behavior near a QCP is to examine how the magnetization and specific heat scale with temperature and magnetic field. This method has been used in studies of $\beta$-YbAlB$_4$,\cite{Matsumoto2011}  YFe$_2$Al$_{10}$,\cite{Wu2014}, CeRu$_4$Sn$_6$,\cite{Fuhrman2021}  and YbAlO$_3$.\cite{Wu2019}  In this paper, we present an extensive scaling analysis of this type for \ybfep. We find scaling behavior that is consistent with the magnetic fluctuations of \ybfep{} having low-dimensional character.

\section{EXPERIMENTAL DETAILS}

Single crystals of \ybfep{} were synthesized from Sn flux.\cite{Asaba2021}  The \ybfep{} single crystals were oriented using a Laue diffractometer in backscattering geometry.  Magnetization measurements were performed in a Quantum Design Magnetic Property Measurement System (MPMS) from 2 K to 40~K and magnetic fields up to 6~T. Specific heat measurements were performed using a Quantum Design Physical Property Measurement System (PPMS) from 0.35 to 20~K and magnetic fields up to 8~T that utilizes a quasi-adiabatic thermal relaxation method.

\section{General Scaling Theory for M(H,T) and C(H,T)}
We consider first how a system scales close to an antiferromagnetic quantum critical point. In the presence of a uniform magnetic field, the scaling is quite distinct from that of a ferromagnet. In ferromagnets, the uniform magnetic field $H$ is conjugate to the ferromagnetic order parameter (the magnetization) and the applied field destroys the phase transition, whereas in an antiferromagnet the applied uniform magnetic field shifts the quantum critical point. In Fig. S1 of the Supplemental Material,\cite{SM} we show a schematic phase diagram of the fixed points and renormalization group (RG) flows of an antiferromagnet in an external uniform magnetic field. For simplicity, we do not consider any tricritical behavior. Notice that the RG flows in Fig. S1 imply that all finite field transitions at T = 0 are governed by the AFQCP fixed point (point 1 in Fig. S1). On the other hand, thermal phase transitions are governed by the thermal fixed point 3 at $T_N$. The $T=0$ critical field $H_c$ (point 2) represents a fixed point for ferromagnetic polarization of the antiferromagnet. This phase diagram provides the basis for the more general scaling analysis presented below. 

Next, we consider a version of the scaling problem that is not specifically tied to antiferromagnetism or ferromagnetism. (Further details are described in the Supplemental Material.\cite{SM}) At the QCP, the free energy is a function of the scaled variables $T/g^\theta$ and $H/g^\phi$, where $g$ is the distance of the driving parameter from the critical value, e.g. $g$ is $P-P_c$ ($y-y_c$) for a pressure (alloy) driven quantum phase transition.  The free energy has the form
\begin{equation}
    f = g^{2-\alpha} F\left[\frac{T}{g^\theta},\frac{H}{g^\phi} \right] 
\label{FE}
\end{equation}	
where $\alpha$ is the free energy exponent.\cite{Ma1985}  We have assumed that the uniform magnetic field is a relevant perturbation at the QCP that scales as $H^{\prime} = b^{z_h}H$ with an exponent $z_h$ that differs from the dynamical exponent $z$ that controls the time scaling. The field exponent $\phi$, which may thus be expressed as $\phi= \nu z_h$, was first introduced  in theoretical studies of disordered random field antiferromagnets,\cite{Ferreira1991,Fishman1979} where a uniform (as opposed to a staggered) magnetic field becomes a relevant parameter. We point out that if $\theta = \phi$, both field and temperature scale the same way, so that the free energy will exhibit $T/H$ scaling. Equation \ref{FE} may be written in the form
	\begin{equation}
    f = H^{\frac{2-\alpha}{\phi}} F\left[\frac{T}{H^{\frac{\theta}{\phi}}}\right]. 
\label{FEn2}
\end{equation}	
Under these assumptions, the magnetization has the form
\begin{equation}
    M = \frac{\partial f}{\partial H} =  H^{\frac{2-\alpha -\phi}{\phi}} F_1\left[\frac{T}{H^{\frac{\theta}{\phi}}}\right].
\label{Magn}
\end{equation}	
and the uniform magnetic susceptibility, which in systems other than ferromagnets will differ from the order parameter susceptibility, has the form
\begin{equation}
    \chi = \frac{\partial^2f}{\partial H^2} = H^{\frac{2-\alpha -2\phi}{\phi}} F_2\left[\frac{T}{H^{\frac{\theta}{\phi}}}\right]	
\label{Chi}
\end{equation}	
and
\begin{equation}
    \frac{d\chi}{dT} = H^{\frac{2-\alpha -\theta -2\phi}{\phi}} F_3\left[\frac{T}{H^{\frac{\theta}{\phi}}}\right]. 	
\label{dchidT}
\end{equation}	
The specific heat coefficient is
\begin{equation}
    \frac{C}{T} = -\frac{\partial^2f}{\partial T^2} = H^{\frac{2-\alpha -2\theta}{\phi}} F_4\left[\frac{T}{H^{\frac{\theta}{\phi}}}\right]. 	
\label{CovT}
\end{equation}	
Assuming that the paramagnetic (non-ordered) state is achieved for $g>0$, we define a field-dependent coherence temperature as
\begin{equation}
    T_{coh} = [g_0 - H^{\frac{1}{\phi}}]^\theta,  	
\label{Tcoh}
\end{equation}	
which marks the onset of Fermi liquid behavior. Similarly, close to an AFQCP, the zero-temperature, field-dependent critical line is given by: $H_c = g^{\phi}$. There are two regimes to consider: In the hyperscaling regime for $d+z\leq4$, the exponent $\theta=\nu z$, where $\nu$ is the exponent for the coherence length as a function of $g$, i.e., $\xi \sim |g|^{-\nu}$.  In addition, the hyperscaling relation holds, $2-\alpha = \nu(d+z)$. In the mean-field/Gaussian regime for $d+z\geq4$, Millis\cite{Millis93} introduced an exponent $\psi$ to characterize the crossover from a Fermi liquid ground state to a regime of classical Gaussian fluctuations, i.e. $T_{FL} \sim |g|^\psi$. In this case, the exponent $\theta = \psi= \frac{z}{d+z-2}$.  In this regime, the exponent $\nu_T$, characterizing the coherence length as a function of temperature, may be different, i.e., $\nu_T \neq \nu$. The $d_{eff}\geq4$ case has both a Gaussian contribution, valid for $H=0$ and where $T_{coh} = |g|^{\theta}$, and a mean-field contribution. In the absence of a magnetic field, the mean-field case shows no fluctuations in the disordered regime, and Gaussian behavior is expected.  It is shown in the Supplemental Material\cite{SM} that in the presence of a magnetic field, the extended mean-field behavior dominates over the Gaussian behavior.

\section{RESULTS}
\subsection{Physical Properties}
\begin{figure*}[t]
\centerline{\includegraphics[width=\linewidth]{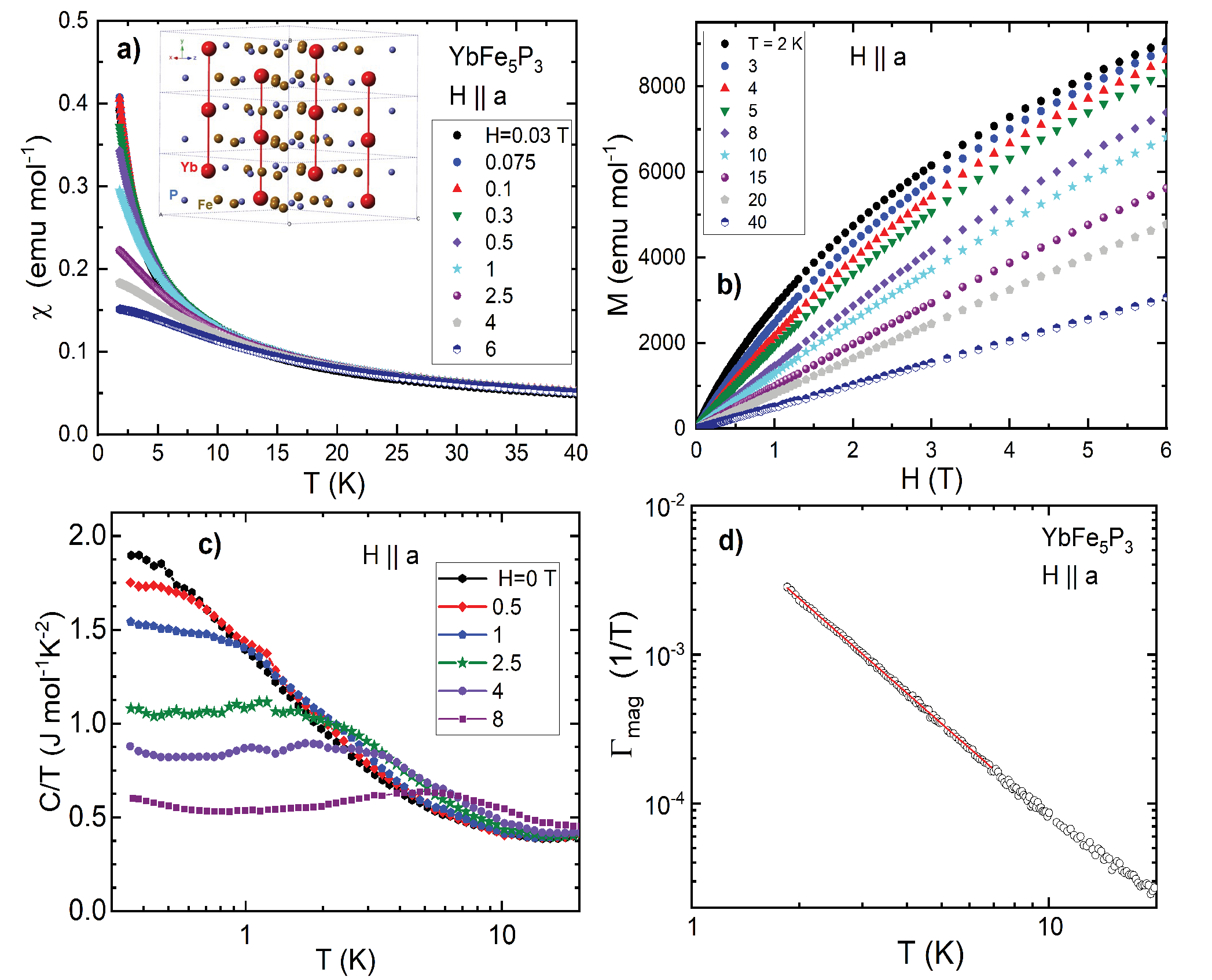}}
\caption{(Color online). a) Magnetic susceptibility $\chi$ vs temperature T of \ybfep{} in various magnetic fields up to $H=6$ T for $H\parallel a$. b) Magnetization $M(H)$ at various temperatures between 2~K and 40~K with $H\parallel a$. c) Specific heat coefficient $C/T$ vs T on a semi-log scale in various applied magnetic fields with $H\parallel a$. d) Magnetic Gr\"{u}neisen parameter $\Gamma_{mag}$ vs T.  The solid line is a fit to the data of the form  $\Gamma_{mag} = BT^{-n}$, yielding $n$ = 2.1.  See text for details.}
\label{props}
\end{figure*}

The magnetic susceptibility $\chi$(T) of \ybfep{} below 40 K in various applied magnetic fields with $H\parallel a$ is shown in Fig. \ref{props}a.  At fields below $\sim$ 0.5 T, $\chi$(T) shows a strong temperature dependence below 10 K, consistent with strong quantum fluctuations as observed previously.\cite{Asaba2021} Above 0.5 T, the magnitude of the magnetic susceptibility is suppressed with increasing field and saturates at the highest field of 6 T. The magnetization $M(H)$ at various temperatures with $H\parallel a$ is displayed in Fig. \ref{props}b.  The $M(H)$ curves have significant curvature for $T<10$ K and are linear in field at higher temperatures up to 40 K.   The specific heat, plotted as $C/T$ vs $T$ on a semi-log scale, in various applied magnetic fields up to 8 T for $H\parallel a$ is shown in Fig. \ref{props}c. For H = 0 T, $C/T \sim -$ln(T) below 10 K and is constant for $T\leq$0.5 K, in agreement with previous results indicating \ybfep{} is close to a quantum critical point.\cite{Asaba2021}  The crossover from a quantum critical regime (for $T\gg H$), where $C/T \sim -$ln(T), and a Fermi-liquid, field polarized regime ($T\ll H^{\frac{\theta}{\phi}}$), in which $C/T \sim$ const., moves to higher temperature with increasing magnetic field (Fig. \ref{props}c).  The suppression of the quantum critical fluctuations in $C(T)/T$ with applied magnetic field is similar to what is found in $\chi(T)$ (Fig. \ref{props}a).  

To further characterize the non-Fermi liquid behavior, the magnetic Gr\"{u}neisen parameter $\Gamma_{mag}(T)$, was determined for \ybfep. The magnetic Gr\"{u}neisen parameter is the ratio of $\partial M/ \partial T$ to specific heat, i.e., $\Gamma_{mag}(T)= -(\partial M/ \partial T) /C_p$ and is expected to diverge at a quantum critical point with magnetic field as the control parameter.\cite{Garst2005,vonLohneysen07,Tokiwa09}  The magnetic Gr\"{u}neisen parameter has been found to diverge in a number of field-tuned quantum critical systems, such as YbRh$_2$Si$_2$,\cite{Tokiwa09} CeCu$_{6-x}$Au$_x$,\cite{vonLohneysen07}, CeRu$_4$Sn$_6$,\cite{Fuhrman2021} and YFe$_{2}$Al$_{10}$.\cite{Wu2014} Figure \ref{props}d shows $\Gamma_{mag}(T)$ of \ybfep{} on a log-log plot in which the data were obtained in a small field of 0.03 T. A fit of a  powerlaw T-dependence from $1.85$ K$\leq T \leq$ 6.9 K, $\Gamma_{mag} = BT^{-n}$, yields $B=0.010$ (K$^{2.1}$T$^{-1}$) and $n$ = 2.1. $\Gamma_{mag}$(T) tends to saturate with the application of larger magnetic fields, as displayed in Fig. S2. These results provide evidence for close proximity to a QCP at $H=0$ for \ybfep, consistent with the other physical properties (Fig. \ref{props}). 

\subsection{Scaling Analysis of \ybfep}
In the hyperscaling regime for $d+z\leq 4$, where $\theta=\nu z$ and $(2-\alpha) = \nu(d+z)$, the experimental scaling data is analyzed in the form (from Eqs.  \ref{Magn}, \ref{dchidT}, and \ref{CovT}):
\begin{equation}
    M\cdot H^{1- \frac{\nu(d+z)}{\phi}} =  F_1\left[\frac{T}{H^{\frac{\theta}{\phi}}}\right] 
\label{MagnExp}
\end{equation}	
\begin{equation}
    \frac{d\chi}{dT} \cdot H^{2- \frac{\nu d}{\phi}} = F_3\left[\frac{T}{H^{\frac{\theta}{\phi}}}\right] 	
\label{dchidTExp}
\end{equation}	
\begin{equation}
    \frac{C}{T} \cdot H^{\frac{\nu(d-z)}{\phi}}=  F_4\left[\frac{T}{H^{\frac{\theta}{\phi}}}\right].
\label{CovTExp}
\end{equation}	

\begin{figure*}[ht]
\centerline{\includegraphics[width=\linewidth]{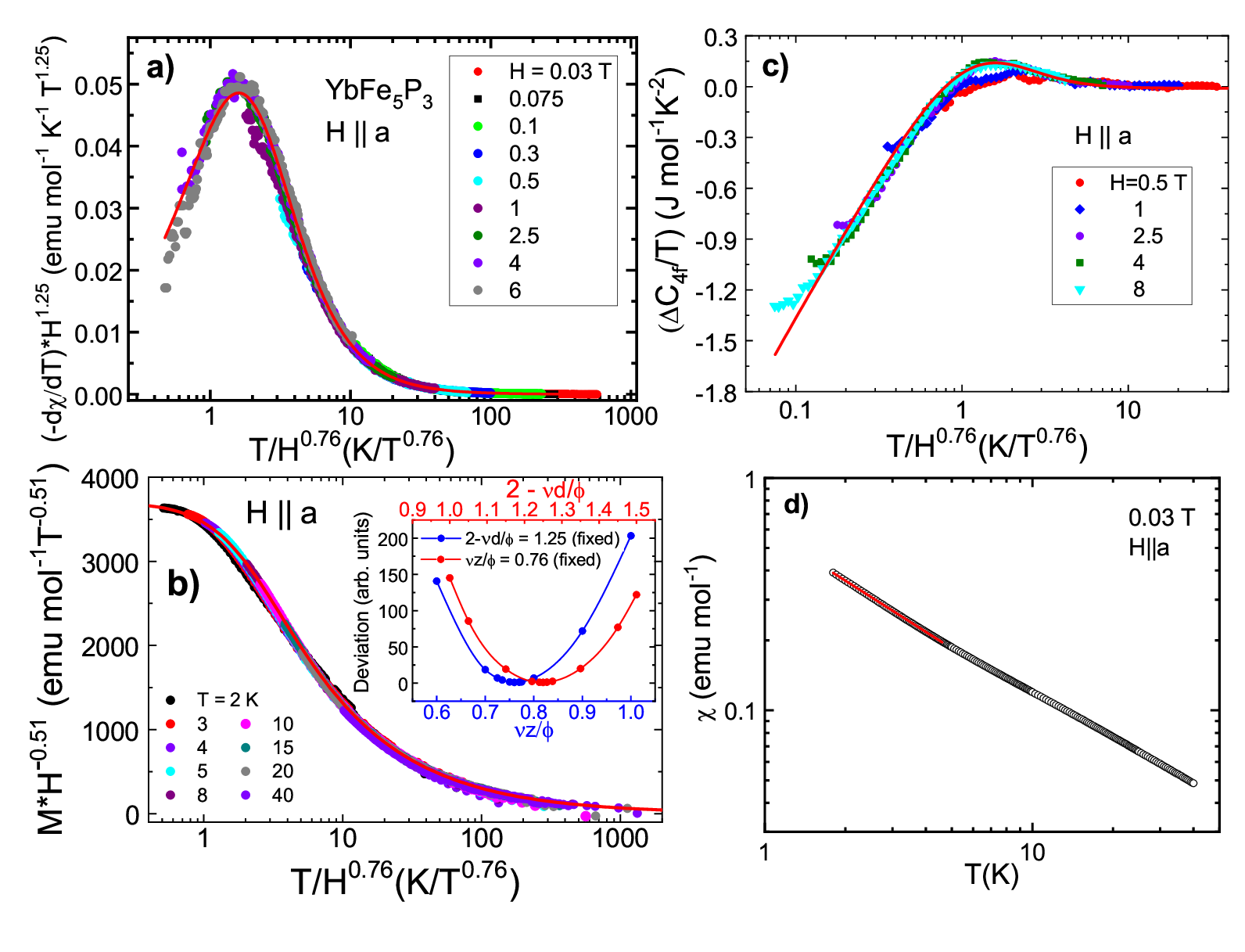}}
\caption{(Color online).  a) Scaling collapse of d$\chi$/dT vs $T/H^{0.76}$ of \ybfep{} in various applied magnetic fields up to 6 T with $H\parallel a$. b) Scaling collapse of $M$ vs $T/H^{0.76}$ at temperatures between 2 and 40 K with $H\parallel a$. The inset of b) shows the fit deviation of the scaling parameters $\nu z/\phi$ with $2-\nu d/\phi$ =1.25 fixed in the scaling analysis (blue curve), while the red curve shows the fit deviation of $2-\nu d/\phi$  with $\nu z/\phi$=0.76 fixed. c) Scaling collapse of $4f$ contribution to the specific heat, $\Delta C_{4f}/T$ vs $T/H^{0.76}$, in various applied magnetic fields with $H\parallel a$. The red lines are fits to the data by the simple expressions for the scaling functions, as discussed in the text. d) Powerlaw fit to $\chi(T)=T^{-\gamma}$ yielding an exponent $\gamma=0.73$, as discussed in the text.   }
\label{scaling}
\end{figure*}

In Fig. \ref{scaling}, we present the scaling analysis of the $d\chi(H, T)/dT$, $M(H,T)$ and $C(H,T)/T$ data of \ybfep{} for $H\parallel a$. (The scaling results for $H\parallel [201]$ and  $H\parallel b$, presented in Figs. S3, and S4, respectively, are similar to the results obtained below for    $H\parallel a$.)  $\Delta C_{4f}(H, T)/T$ is defined as $C_{4f}(H)/T - C_{4f}(0)/T$, where $C_{4f}/T$ is the $4f$ contribution to C/T after subtraction of the nonmagnetic LuFe$_5$P$_3$ contribution.\cite{Asaba2021} 
From Figs. \ref{scaling}a,b,c and Eqs.  \ref{MagnExp}, \ref{dchidTExp},  and \ref{CovTExp}, we find that: 
	\begin{align}
        & \nu z/\phi=0.76,\nonumber\\ 
   & 2- \frac{\nu d}{\phi}=1.25, \rightarrow  \frac{\nu d}{\phi}=0.75,\nonumber\\ 
      & 1 - \frac{\nu(z+d)}{\phi} = -0.51 \nonumber\\
    &\frac{\nu(z-d)}{\phi}=0.01
    \label{exponents}
\end{align}
 The deviation of the scaling, shown in the inset of Fig. \ref{scaling}b, places stringent limits on the scaling parameters. The main result for \ybfep{} in the hyperscaling regime is that $d=z$. If we associate the logarithmic behavior of the specific heat displayed in Fig. \ref{props}c with a value $\alpha=0$ for the free energy exponent, then we have $\nu (d+z)=2$, so that $\nu z = \nu d =1$; hence, $\phi=4/3$.  At the QCP, the magnetic susceptibility has the form (in the limit of small fields):
\begin{equation}
    \chi = T^{-\gamma}, 
\label{chi2}
\end{equation}	
where $\gamma= \frac{2\phi   -(2-\alpha)}{\theta}$.
These values imply that the exponent of the uniform susceptibility has the value:
\begin{equation}
     \gamma= \frac{2\phi}{\nu z} -  \frac{(d+z)}{z} =2/3.
\label{gamma}
\end{equation}	
In a small magnetic field H=0.03 T, $\chi(T)$ exhibits a powerlaw temperature dependence as shown in Fig. \ref{scaling}d, albeit over a small temperature range 1.8 K $<$ T $<$ 4 K and yields the observed value $\gamma = 0.73$, close to the expected value 2/3. 

Following the scaling results for  YFe$_2$Al$_{10}$ (Ref. \cite{Wu2014}) and YbAlB$_4$,\cite{Matsumoto2011} a simple expression for the scaling function of the magnetization (Eq. (\ref{Magn})) is given by: 
\begin{equation}
M =   H^{\frac{2-\alpha -\phi}{\phi}} F_1(x) \nonumber
\end{equation}	
\begin{equation}
\mathrm{with} \,\,\, F_1(x) = c(x^2+a^2)^{-\gamma/2}, 
\label{MagnSimple}
\end{equation}	
where $x=T/H^{\theta/\phi}$ and $c$ and $a$ are constants. This leads to an explicit expression for the scaling function for $d\chi/dT$:
\begin{equation}
\frac{d\chi}{dT} = H^{\frac{2-\alpha -\theta -2\phi}{\phi}} F_3(x)
 \nonumber
\end{equation}
\begin{equation}
 \mathrm{with} \,\, F_3(x) = -c\gamma x (x^2+a^2)^{-\gamma/2-1}.
\label{dchidTSimple}
\end{equation}
A fit of the scaling function in Eq. (\ref{dchidTSimple}) to the scaled $d\chi/dT$ data, using the experimental values obtained from the scaling analysis of Eqs. \ref{MagnExp}, \ref{dchidTExp}, and \ref{CovTExp}, yields $a = 2.05$ and $c=0.59$.  As displayed in Figs. \ref{scaling}a and \ref{scaling}b, the agreement between the proposed scaling function (red lines) and the scaled data is excellent over several decades in $x$. The result $d=z$ derived from the scaling analysis leads to a simple form for the field-dependent specific heat:  $\Delta C_{4f}(H, T)/T = H^0\, F_4(x)$, where $F_4(x)$ is related to $F_1(x)$ by the Maxwell relation $\partial S/\partial H = \partial M/\partial T$. The scaling expression of $\Delta C_{4f}/T$ derived from the Maxwell relation is shown in Fig. \ref{scaling}c and is in good agreement with the $\Delta C_{4f}(H, T)/T$ data (see  Ref. \cite{SM} for details). The self-consistency among the scaling forms for $M$, $d\chi/dT$, $\Delta C_{4f}/T$ provide support for the robustness of the scaling analysis of \ybfep. 

The temperature and magnetic field dependence the magnetic Gr\"{u}neisen parameter of \ybfep{} is predicted to be $\Gamma_{mag} \sim H^{\frac{\theta}{\phi}-1} F_1^{\prime}(x)/F_4(x)$ from the scaling analysis (Eqs. \ref{Magn} and \ref{CovT}). This prediction yields a $\sim T^{-2}$ temperature dependence, consistent with experiment (Fig. \ref{props}d). (Explicitly, $\Gamma_{mag} \sim T^{-8/3}/(-ln(T/T_K)$ with $T_K=$8 K from Eq. (\ref{MagnSimple}) (for $x\gg a$) and Fig. \ref{props}a, as described in the Supplemental Material,\cite{SM} which is in agreement with the observed powerlaw T-dependence).  However, the field-dependent scaling, which is predicted to be $\Gamma_{mag} H = f(T/H^{0.76})$, is only observed for $H\geq 2.5$ T (not shown).  
     
\begin{figure*}[ht]
\centerline{\includegraphics[width=\linewidth]{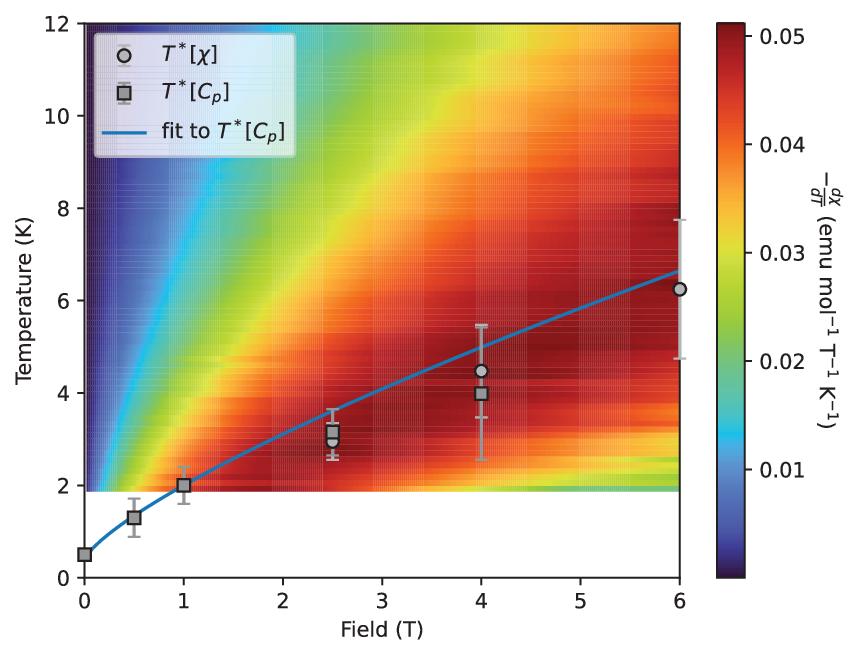}}
\caption{(Color online).  Temperature-Magnetic Field ($T-H$) phase diagram of \ybfep.  The $T-H$ contour plot of $d\chi/dT$ shows the crossover from quantum critical to a Fermi liquid state.  The  squares (circles) represent $T^\star (H)$, taken from the maximum in the scaled $\Delta C_{4f}/T$  (scaled $d\chi/dT$ data) at various magnetic fields (see Fig. \ref{scaling}c). The light blue line is a fit to the data to Eq. (\ref{Tstar}), as discussed in the text.  }
\label{phase}
\end{figure*}

In Fig. \ref{phase}, we plot the phase diagram for \ybfep{} as a function of temperature and magnetic field based upon our scaling results. The color plot in the $H-T$ plane represents the magnitude of the scaled $d\chi/dT$ data from Fig. \ref{scaling}a.  The crossover temperature is defined as $T^\star = x^\star H^{0.76}$, shown in Fig. \ref{phase}, is determined from the maxima in the scaled $d\chi/dT$ and $\Delta C_{4f}/T$ data (see Figs. \ref{scaling}a,c); $T^\star(H)$ represents the crossover from the low-dimensional quantum critical (QC) state (red region) to a 3D Fermi liquid state (blue region), which we believe to be the ultimate ground state. As pointed out above, fluctuations in a 1D or 2D system cross over to a Fermi liquid  ground state with renormalized parameters for $T\ll T^\star(H)$. The equation for this line in the case of \ybfep, which is not exactly at the QCP ($g>0$), is given by: 
\begin{equation}
    T^\star = [T_0 - x^\star H^{\frac{1}{\phi}}]^\theta  	
\label{Tstar}
\end{equation}	
Figure  \ref{phase} shows a fit of this equation, with $\theta=\nu z = 1$ (fixed), $\phi=4/3$ (fixed), and $T_0=0.47$~K and $x^\star$=1.63.  The fit is in good agreement with the experimental data and with the Fermi liquid state below 0.5 K (Fig. \ref{props}c). In Fig. S5, we present two possible scenarios for an AFQCP, accessed either by pressure or chemical substitution.  Previous measurements of \ybfep{} under pressure indicate a QCP at a critical pressure $P_c\sim 3$ GPa, with an extended range of $T$-linear electrical resistivity.\cite{Asaba2021}   While pressure is a clean tuning variable, which does not introduce disorder, chemical substitution may significantly influence the nature of the QCP.  Indeed, preliminary measurements on Yb(Fe$_{1-y}$Co$_y$)$_5$P$_3$, in which Co compresses the lattice, reveals $H/T$ scaling at a critical concentration $y_c\sim 0.06$; for $y> 0.06$, antiferromagnetic order appears with $\mathbf{Q_N}$=(0,0,0) AFM state.\cite{Avers2024}

\section{DISCUSSION}
The scaling results for \ybfep{} narrow the down the possible types of quantum fluctuations that arise near the QCP.   First, we consider the mean-field / Gaussian regime for $d+z \geq$ 4: as $\psi/\phi \approx $3/4 and $(\psi + 2\phi -2)/\phi \approx 5/4$ (assuming the free energy exponent $\alpha =0$ in mean-field), we must have $\phi =4/3$ and $\psi =1$. The expression for $\psi =z/(d+z-2)$ will automatically be satisfied for $d=2$ and for any value of $z$. Assuming $z=3$ represents ferromagnetic behavior\cite{vonLohneysen07}, the case $d=2$, $z=3$ is ruled out because $\phi =\beta + \gamma =3/2$  for a mean-field ferromagnet as described in the Supplemental Material\cite{SM}, which is contrary to our result $\phi=4/3$.  The scaling results for \ybfep{} are appropriate for a two-dimensional quantum critical antiferromagnet with $d=z=2$ at the upper critical dimension, $d_{eff} =d+z=4$. The $-log(T)$ dependence of the specific heat coefficient is in agreement with prediction for an AFQCP with $d=z=2$. However, the Yb chains are separated by distances greater than or equal to 5.61 \AA{} so that when viewed down the chain b-axis, 2D layers of Yb atoms can be seen with inter-chain distances at least 1.5 times greater than the intra-chain distance. This suggests that the layers would be coupled more strongly along the chain direction than within the planes, which would tend to create a pseudo-1D magnetic environment. We  rule out that the fluctuations are three dimensional because for $d=3$, we have $\psi=z/(z+1)\neq1$.  If the fluctuations are indeed 1D, as the presence of Yb chains along the b-axis in \ybfep{} and the predicted 1D Fermi surface sheets in \ybfep{} suggest,\cite{Asaba2021}  then in the hyperscaling regime for $d+z\leq 4$ (as discussed above),  the scaling conditions lead to the conclusion that $z=1$ and also $\nu =1$. A dynamic exponent $z=1$ would be unusual for an antiferromagnet, but according to Millis,\cite{Millis93} this can occur if the critical fluctuations lie outside of the particle-hole continuum. For the 1D system YbAlO$_3$, this value of the dynamic exponent $z=1$, which has been observed in the vicinity of the polarization transition in high magnetic field, has been attributed to exchange anisotropy.\cite{Wu2019}  Reference  \cite{Danu2022} points out that a 1D $S=1/2$ Heisenberg chain coupled by a Kondo interaction to a 2D-electron gas exhibits a crossover from $z=1$ to $z=2$ as the AFQCP is approached, so that a region of $d=1, z=1$ behavior is possible. Physically, the relevance of the uniform magnetic at the zero temperature Kondo-lattice fixed point is related to a competition between  fully screened and magnetic polarized ground states.

In both cases presented for \ybfep, we expect $E/T$ scaling of the dynamic susceptibility:  For the hyperscaling case ($d=z=\nu=1$), the energy scale varies as $E \sim T^{\nu z} \sim T$. For the mean-field/Gaussian case ($d=z=2$),  where $\nu_T=\nu /\psi$, $E \sim T^{\nu_T z} \sim T$, because $\psi=1$.  We interpret the $T-$linear resistivity seen above 0.5 K as non-Fermi liquid behavior associated with low-dimensionality in \ybfep. However, the rollover towards Fermi liquid $T^2$ behavior at lower temperatures than 0.5 K at ambient pressure suggests a crossover to 3D physics at the lowest temperatures. This 3D state might be the precursor of the AFM structure seen in Yb(Fe$_{1-y}$Co$_y$)$_5$P$_3$,\cite{Avers2024} which is clearly 3D. Because the scaling behavior that we have observed was determined at temperatures above 0.35 K, the implied low-dimensionality would not be affected by the crossover into a 3D Fermi liquid state. Hence, the 1D quantum critical behavior must be confined to the spin chains, with weak interchain coupling above $T^\star$ 0.5 K. Importantly, the fact that the material appears to enter a Fermi liquid phase below 0.5 K also implies that the material does not become a Luttinger liquid.\cite{Giamarchi2003,Birgenau1978} Therefore, the scaling results leave only two scenarios: If the fluctuations are one-dimensional, then $\nu=z=1$. If they are two-dimensional, then under hyperscaling (valid at $z=d=2$, so that $d+z=4$) in order to satisfy $\nu z=1$, we must have $\nu =1/2$, the mean-field exponent. We conclude that the fluctuations responsible for the scaling behavior are in reduced dimensions ($d < 3$).  

In most $f$-electron materials where the scaling of the magnetization and specific heat has been measured as a function of both temperature and magnetic field, $T/H$ scaling is observed. These materials include the oxide YbAlO$_3$,\cite{Wu2019}  the mixed-valent metal $\beta$-YbAlB$_4$,\cite{Matsumoto2011} and the heavy fermion compounds CeCu$_{6-x}$Au$_x$ and YbRh$_2$(Ge$_{1-x}$Si$_{x}$)$_2$.\cite{Schroder2000,Custers03}  We have also observed $T/H$ scaling at the QCP in alloys of \ybfep{} with Co and Ru.\cite{Avers2024} Two exceptions to $T/H$ scaling are YFe$_2$Al$_{10}$ (Ref. \cite{Wu2014}) and CeRu$_4$Sn$_6$.\cite{Fuhrman2021} In YFe$_2$Al$_{10}$,  the values $\nu z/\phi =3/5$ and $2-\nu d/\phi =7/5$ observed when Eqs. \ref{dchidT} and \ref{CovT} are applied,  also satisfy the condition $z=d$ that is found in \ybfep. Other than the difference that $\phi$=5/3, the same scaling analysis of YFe$_2$Al$_{10}$ yields identical results: $\psi$=1 and the possible scenarios are restricted to $d=z=\nu=1$, and $d=z=2$ with $\nu=1/2$. The orthorhombic YFe$_2$Al$_{10}$ structure is layered and, hence, possibly two-dimensional, consistent with $z=d=2$.  Magnetic dynamics exhibit $E/T$ scaling in this system,\cite{Gannon2018} as expected for the $\nu=1/2$, $z=2$ case. The measured exponent for the uniform susceptibility is $\gamma =1.4$, close to the value 4/3 expected from the Eq. (\ref{gamma}).  Originally for YFe$_2$Al$_{10}$,\cite{Wu2014} it was assumed that this exponent meant that the material is near to a ferromagnetic QCP, but a diverging uniform susceptibility is possible close to an antiferromagnetic QCP (see Supplemental Material\cite{SM}), and the value of the exponent $\phi=5/3$ is not equal to the expected value $\phi=\beta + \gamma=3/2$ for a 2D ferromagnetic QCP.\cite{Continentino2017} In subsequent inelastic neutron scattering measurements of YFe$_2$Al$_{10}$,\cite{Gannon2018} fluctuations of the Fe moments were found to have no $Q$-dependence beyond that of the calculated form factor. This lack of Q-dependence implies that the magnetic fluctuations are local. A problem that arises with scenario is that for fluctuations that satisfy hyperscaling, Eq. (\ref{FEn2}) takes the form $f \sim H^{\nu (d+z)/\phi}\,\, F(T/H^{\nu z/\phi}$). For $d=0$, this implies that the exponent of H of the left-hand side of Eqns. \ref{MagnExp} and \ref{dchidTExp}  has the same value as $\theta/\phi$ in $F_3$ and $F_4$, in violation of the experimental scaling results. In addition, the scaling analysis, which gives $d=1$ or 2, is also inconsistent with $d=0$ fluctuations. The values of $d=z=1$ or 2 for these two cases are also contrary to the expectation that the dispersion of either magnons or the overdamped critical fluctuations should have the form $\omega \sim (Q-Q_N)^z$, where $Q_N$ is the ordering wavevector.\cite{vonLohneysen07}  We note that the authors of Ref. \cite{Gannon2018} argue that the quantum transition is not associated with suppression of a magnetic transition, but instead results from an orbital selective Mott transition,\cite{Pepin2007,Vojta2010,deMedici2005} which may require a different scaling analysis. It is often the case for heavy Fermion compounds near a QCP that experimentally observed scaling behavior may correspond to a fluctuation that does not ultimately become critical. An example of this situation is the  CeCu$_{6-x}$Ag$_x$ system,\cite{Poudel2019} where the scaling law is dominated by a $Q$=(1,0,0) fluctuation that exhibits $\omega/T$ scaling, but the actual critical point occurs for $Q_N=(0.65,0,0.3)$, for which the scaling is of the form $\omega/T^{3/2}$ expected for Hertz-Millis-Moriya systems.\cite{vonLohneysen07}  However, because the experimentally observed scaling laws reflect the behavior of the Fe magnetization in YFe$_2$Al$_{10}$,\cite{Gannon2018} and the observation of a local fluctuations of the Fe atoms appears to violate the scaling behavior,\cite{Wu2014} it is a current dilemma how to reconcile the observed scaling parameters with the zero-dimensional magnetic fluctuations and the $E/T$ scaling in YFe$_2$Al$_{10}$.\cite{Wu2014,Gannon2018} The tetragonal compound CeRu$_4$Sn$_6$ is another example of a material that does not exhibit $H/T$ scaling.\cite{Fuhrman2021}  In the Supplemental Material, we apply Eq. (\ref{dchidTExp}) to the data to show that in the hyperscaling regime, small values of $z\sim 0.3-0.8$ are obtained for $d=1,2,$ or 3.  These values are quite unusual.  The authors interpret the neutron scattering results as a signature of local Kondo breakdown criticality.\cite{Fuhrman2021}  This leads to contradictions between the scaling results and the expectations for the behavior of a local ($d=0$) system, similar to those discussed above for YFe$_2$Al$_{10}$.

Without neutron scattering results for \ybfep, it is not known whether this compound also has localized fluctuations as observed in YFe$_2$Al$_{10}$. Nevertheless, it is possible that a more unusual scenario than an AFQCP may  apply to \ybfep, such as the suggested selective orbital Mott transition proposed for YFe$_2$Al$_{10}$.\cite{Gannon2018} While our general scaling analysis yields the exponents $z, \nu, \psi$, and $\phi$, without microscopic verification, the nature of the QCP remains “undisclosed.”

\section{CONCLUSIONS}

We measured the low temperature behavior of the magnetization and specific heat of the quasi-one dimensional spin chain system \ybfep. We compared the behavior to the predictions of scaling laws that are based on very general considerations about the temperature and magnetic field dependence of an antiferromagnetic system close to a quantum critical point. In this approach, the free energy depends on the parameters $d$, $\nu$, and $z$. For the case where the uniform magnetic field is a relevant operator for an AFQCP, a fourth exponent $\phi$ is included in the free energy; this exponent is related to a new dynamic exponent $z_h$ via $\phi=\nu z_h$. When $z\neq z_h$, it results in scaling behavior that does not depend simply on $H/T$.  An advantage of this approach is that it can limit the behavior to a restricted set of values of the above parameters. We have found excellent agreement between the data and the scaling laws and find that for \ybfep, there are only two possible cases: $d=1, \nu=1, z=1$, and $d=2$, $\nu=1/2$, and $z=2$, both with $\phi =4/3$. While this approach successfully determines these exponents, and restricts the possible underlying physics, it cannot determine the latter without some form of microscopic verification, such as inelastic neutron scattering. In systems such as YFe$_2$Al$_{10}$ and CeRu$_4$Sn$_6$ where critical scaling is observed that is distinct from "$H/T$" scaling, neutron scattering indicates $\omega/T$ scaling that is interpreted as being associated with an orbital-selective Mott transition or a local QCP.\cite{Gannon2018,Fuhrman2021}  It is currently a conundrum of how to relate the thermodynamic scaling results to the neutron scattering in these materials. We hope that results in the quantum critical system \ybfep, as well as YFe$_2$Al$_{10}$ and CeRu$_4$Sn$_6$, will stimulate further theoretical and experimental work to understand the unusual critical scaling of these materials.

\section{ACKNOWLEDGMENTS}

We thank C. Batista, A. Dioguardi, M. Kenzelmann, A. Scheie, and Q. Si for helpful discussions. Work at Los Alamos National Laboratory was performed under the auspices of the U.S. Department of Energy, Office of Basic Energy Sciences, Division of Materials Science and Engineering under project ``Quantum Fluctuations in Narrow-Band Systems". M.A.C. acknowledges partial financial support from FAPERJ, Grant E-26/201.223/2021, and CNPq (Grant 305810/2020-0).

\bibliography{YbFe5P3_bib}

\begin{thebibliography}{13}%
\makeatletter
\providecommand \@ifxundefined [1]{%
 \@ifx{#1\undefined}
}%
\providecommand \@ifnum [1]{%
 \ifnum #1\expandafter \@firstoftwo
 \else \expandafter \@secondoftwo
 \fi
}%
\providecommand \@ifx [1]{%
 \ifx #1\expandafter \@firstoftwo
 \else \expandafter \@secondoftwo
 \fi
}%
\providecommand \natexlab [1]{#1}%
\providecommand \enquote  [1]{``#1''}%
\providecommand \bibnamefont  [1]{#1}%
\providecommand \bibfnamefont [1]{#1}%
\providecommand \citenamefont [1]{#1}%
\providecommand \href@noop [0]{\@secondoftwo}%
\providecommand \href [0]{\begingroup \@sanitize@url \@href}%
\providecommand \@href[1]{\@@startlink{#1}\@@href}%
\providecommand \@@href[1]{\endgroup#1\@@endlink}%
\providecommand \@sanitize@url [0]{\catcode `\\12\catcode `\$12\catcode
  `\&12\catcode `\#12\catcode `\^12\catcode `\_12\catcode `\%12\relax}%
\providecommand \@@startlink[1]{}%
\providecommand \@@endlink[0]{}%
\providecommand \url  [0]{\begingroup\@sanitize@url \@url }%
\providecommand \@url [1]{\endgroup\@href {#1}{\urlprefix }}%
\providecommand \urlprefix  [0]{URL }%
\providecommand \Eprint [0]{\href }%
\providecommand \doibase [0]{https://doi.org/}%
\providecommand \selectlanguage [0]{\@gobble}%
\providecommand \bibinfo  [0]{\@secondoftwo}%
\providecommand \bibfield  [0]{\@secondoftwo}%
\providecommand \translation [1]{[#1]}%
\providecommand \BibitemOpen [0]{}%
\providecommand \bibitemStop [0]{}%
\providecommand \bibitemNoStop [0]{.\EOS\space}%
\providecommand \EOS [0]{\spacefactor3000\relax}%
\providecommand \BibitemShut  [1]{\csname bibitem#1\endcsname}%
\let\auto@bib@innerbib\@empty
\bibitem [{\citenamefont {Millis}(1993)}]{Millis93}%
  \BibitemOpen
  \bibfield  {author} {\bibinfo {author} {\bibfnamefont {A.~J.}\ \bibnamefont
  {Millis}},\ }\href@noop {} {\bibfield  {journal} {\bibinfo  {journal} {Phys.
  Rev. B}\ }\textbf {\bibinfo {volume} {48}},\ \bibinfo {pages} {7183}
  (\bibinfo {year} {1993})}\BibitemShut {NoStop}%
\bibitem [{\citenamefont {Asaba}\ \emph {et~al.}(2021)\citenamefont {Asaba},
  \citenamefont {Lee}, \citenamefont {Seo}, \citenamefont {Avers},
  \citenamefont {Thomas}, \citenamefont {Movshovich}, \citenamefont {Thompson},
  \citenamefont {Rosa}, \citenamefont {Bauer},\ and\ \citenamefont
  {Ronning}}]{Asaba2021}%
  \BibitemOpen
  \bibfield  {author} {\bibinfo {author} {\bibfnamefont {T.}~\bibnamefont
  {Asaba}}, \bibinfo {author} {\bibfnamefont {S.}~\bibnamefont {Lee}}, \bibinfo
  {author} {\bibfnamefont {S.}~\bibnamefont {Seo}}, \bibinfo {author}
  {\bibfnamefont {K.~E.}\ \bibnamefont {Avers}}, \bibinfo {author}
  {\bibfnamefont {S.~M.}\ \bibnamefont {Thomas}}, \bibinfo {author}
  {\bibfnamefont {R.}~\bibnamefont {Movshovich}}, \bibinfo {author}
  {\bibfnamefont {J.~D.}\ \bibnamefont {Thompson}}, \bibinfo {author}
  {\bibfnamefont {P.~F.~S.}\ \bibnamefont {Rosa}}, \bibinfo {author}
  {\bibfnamefont {E.~D.}\ \bibnamefont {Bauer}},\ and\ \bibinfo {author}
  {\bibfnamefont {F.}~\bibnamefont {Ronning}},\ }\href
  {https://doi.org/10.1103/PhysRevB.104.195140} {\bibfield  {journal} {\bibinfo
   {journal} {Phys. Rev. B}\ }\textbf {\bibinfo {volume} {104}},\ \bibinfo
  {pages} {195140} (\bibinfo {year} {2021})}\BibitemShut {NoStop}%
\bibitem [{K.E. Avers et al.()}]{Avers2024}%
  \BibitemOpen
  K.E. Avers et al.,\ \href@noop {} {} (\bibinfo {year} {2024}),\ \bibinfo
  {note} {(unpublished)}\BibitemShut {NoStop}%
\bibitem [{\citenamefont {Wu}\ \emph {et~al.}(2014)\citenamefont {Wu},
  \citenamefont {Kim}, \citenamefont {Park}, \citenamefont {Tsvelik},\ and\
  \citenamefont {Aronson}}]{Wu2014}%
  \BibitemOpen
  \bibfield  {author} {\bibinfo {author} {\bibfnamefont {L.~S.}\ \bibnamefont
  {Wu}}, \bibinfo {author} {\bibfnamefont {M.~S.}\ \bibnamefont {Kim}},
  \bibinfo {author} {\bibfnamefont {K.}~\bibnamefont {Park}}, \bibinfo {author}
  {\bibfnamefont {A.~M.}\ \bibnamefont {Tsvelik}},\ and\ \bibinfo {author}
  {\bibfnamefont {M.~C.}\ \bibnamefont {Aronson}},\ }\href
  {https://www.pnas.org/doi/abs/10.1073/pnas.1413112111} {\bibfield  {journal}
  {\bibinfo  {journal} {Proceedings of the National Academy of Sciences}\
  }\textbf {\bibinfo {volume} {111}},\ \bibinfo {pages} {14088} (\bibinfo
  {year} {2014})}\BibitemShut {NoStop}%
\bibitem [{\citenamefont {Ma}(1985)}]{Ma1985}%
  \BibitemOpen
  \bibfield  {author} {\bibinfo {author} {\bibfnamefont {S.-K.}\ \bibnamefont
  {Ma}},\ }\href {https://doi.org/10.1142/0073} {\emph {\bibinfo {title}
  {Statistical Mechanics}}}\ (\bibinfo  {publisher} {World Scientific},\
  \bibinfo {year} {1985})\ 
 \BibitemShut {NoStop}%
\bibitem [{\citenamefont {Continentino}(2001)}]{Continentino2001}%
  \BibitemOpen
  \bibfield  {author} {\bibinfo {author} {\bibfnamefont {M.~A.}\ \bibnamefont
  {Continentino}},\ }\href {https://doi.org/10.1142/4498} {\emph {\bibinfo
  {title} {Quantum Scaling in Many-Body Systems}}}\ (\bibinfo  {publisher}
  {World Scientific},\ \bibinfo {year} {2001})\ 
  \BibitemShut {NoStop}%
\bibitem [{\citenamefont {Moriya}\ and\ \citenamefont
  {Takimoto}(1995)}]{Moriya95}%
  \BibitemOpen
  \bibfield  {author} {\bibinfo {author} {\bibfnamefont {T.}~\bibnamefont
  {Moriya}}\ and\ \bibinfo {author} {\bibfnamefont {T.}~\bibnamefont
  {Takimoto}},\ }\href@noop {} {\bibfield  {journal} {\bibinfo  {journal} {J.
  Phys. Soc. Japan}\ }\textbf {\bibinfo {volume} {64}},\ \bibinfo {pages} {960}
  (\bibinfo {year} {1995})},\ \bibinfo {note} {and refs. therein}\BibitemShut
  {NoStop}%
\bibitem [{\citenamefont {Ramazashvili}(1999)}]{Ramazashvili1999}%
  \BibitemOpen
  \bibfield  {author} {\bibinfo {author} {\bibfnamefont {R.}~\bibnamefont
  {Ramazashvili}},\ }\href {https://doi.org/10.1103/PhysRevB.60.7314}
  {\bibfield  {journal} {\bibinfo  {journal} {Phys. Rev. B}\ }\textbf {\bibinfo
  {volume} {60}},\ \bibinfo {pages} {7314} (\bibinfo {year}
  {1999})}\BibitemShut {NoStop}%
\bibitem [{\citenamefont {Continentino}(2004)}]{Continentino2004}%
  \BibitemOpen
  \bibfield  {author} {\bibinfo {author} {\bibfnamefont {M.~A.}\ \bibnamefont
  {Continentino}},\ }\href
  {https://doi.org/https://doi.org/10.1016/j.jmmm.2003.11.097} {\bibfield
  {journal} {\bibinfo  {journal} {Journal of Magn. \& Magn. Mater.}\
  }\textbf {\bibinfo {volume} {272-276}},\ \bibinfo {pages} {231} (\bibinfo
  {year} {2004})}\BibitemShut {NoStop}%
\bibitem [{\citenamefont {Matsumoto}\ \emph {et~al.}(2011)\citenamefont
  {Matsumoto}, \citenamefont {Nakatsuji}, \citenamefont {Kuga}, \citenamefont
  {Karaki}, \citenamefont {Horie}, \citenamefont {Shimura}, \citenamefont
  {Sakakibara}, \citenamefont {Nevidomskyy},\ and\ \citenamefont
  {Coleman}}]{Matsumoto2011}%
  \BibitemOpen
  \bibfield  {author} {\bibinfo {author} {\bibfnamefont {Y.}~\bibnamefont
  {Matsumoto}}, \bibinfo {author} {\bibfnamefont {S.}~\bibnamefont
  {Nakatsuji}}, \bibinfo {author} {\bibfnamefont {K.}~\bibnamefont {Kuga}},
  \bibinfo {author} {\bibfnamefont {Y.}~\bibnamefont {Karaki}}, \bibinfo
  {author} {\bibfnamefont {N.}~\bibnamefont {Horie}}, \bibinfo {author}
  {\bibfnamefont {Y.}~\bibnamefont {Shimura}}, \bibinfo {author} {\bibfnamefont
  {T.}~\bibnamefont {Sakakibara}}, \bibinfo {author} {\bibfnamefont {A.~H.}\
  \bibnamefont {Nevidomskyy}},\ and\ \bibinfo {author} {\bibfnamefont
  {P.}~\bibnamefont {Coleman}},\ }\href
  {https://www.science.org/doi/abs/10.1126/science.1197531} {\bibfield
  {journal} {\bibinfo  {journal} {Science}\ }\textbf {\bibinfo {volume}
  {331}},\ \bibinfo {pages} {316} (\bibinfo {year} {2011})}\BibitemShut
  {NoStop}%
\bibitem [{\citenamefont {Fuhrman}\ \emph {et~al.}(2021)\citenamefont
  {Fuhrman}, \citenamefont {Sidorenko}, \citenamefont {Haenel}, \citenamefont
  {Winkler}, \citenamefont {Prokofiev}, \citenamefont {Rodriguez-Rivera},
  \citenamefont {Qiu}, \citenamefont {Blaha}, \citenamefont {Si}, \citenamefont
  {Broholm},\ and\ \citenamefont {Paschen}}]{Fuhrman2021}%
  \BibitemOpen
  \bibfield  {author} {\bibinfo {author} {\bibfnamefont {W.~T.}\ \bibnamefont
  {Fuhrman}}, \bibinfo {author} {\bibfnamefont {A.}~\bibnamefont {Sidorenko}},
  \bibinfo {author} {\bibfnamefont {J.}~\bibnamefont {Haenel}}, \bibinfo
  {author} {\bibfnamefont {H.}~\bibnamefont {Winkler}}, \bibinfo {author}
  {\bibfnamefont {A.}~\bibnamefont {Prokofiev}}, \bibinfo {author}
  {\bibfnamefont {J.~A.}\ \bibnamefont {Rodriguez-Rivera}}, \bibinfo {author}
  {\bibfnamefont {Y.}~\bibnamefont {Qiu}}, \bibinfo {author} {\bibfnamefont
  {P.}~\bibnamefont {Blaha}}, \bibinfo {author} {\bibfnamefont
  {Q.}~\bibnamefont {Si}}, \bibinfo {author} {\bibfnamefont {C.~L.}\
  \bibnamefont {Broholm}},\ and\ \bibinfo {author} {\bibfnamefont
  {S.}~\bibnamefont {Paschen}},\ }\href
  {https://doi.org/10.1126/sciadv.abf9134} {\bibfield  {journal} {\bibinfo
  {journal} {Science Advances}\ }\textbf {\bibinfo {volume} {7}},\ \bibinfo
  {pages} {eabf9134} (\bibinfo {year} {2021})}\BibitemShut {NoStop}%
\bibitem [{\citenamefont {Ferreira}\ \emph {et~al.}(1991)\citenamefont
  {Ferreira}, \citenamefont {King},\ and\ \citenamefont
  {Jaccarino}}]{Ferreira1991}%
  \BibitemOpen
  \bibfield  {author} {\bibinfo {author} {\bibfnamefont {I.~B.}\ \bibnamefont
  {Ferreira}}, \bibinfo {author} {\bibfnamefont {A.~R.}\ \bibnamefont {King}},\
  and\ \bibinfo {author} {\bibfnamefont {V.}~\bibnamefont {Jaccarino}},\ }\href
  {https://doi.org/10.1103/PhysRevB.43.10797} {\bibfield  {journal} {\bibinfo
  {journal} {Phys. Rev. B}\ }\textbf {\bibinfo {volume} {43}},\ \bibinfo
  {pages} {10797} (\bibinfo {year} {1991})}\BibitemShut {NoStop}%
\bibitem [{\citenamefont {Fishman}\ and\ \citenamefont
  {Aharony}(1979)}]{Fishman1979}%
  \BibitemOpen
  \bibfield  {author} {\bibinfo {author} {\bibfnamefont {S.}~\bibnamefont
  {Fishman}}\ and\ \bibinfo {author} {\bibfnamefont {A.}~\bibnamefont
  {Aharony}},\ }\href {https://dx.doi.org/10.1088/0022-3719/12/18/006}
  {\bibfield  {journal} {\bibinfo  {journal} {Journal of Physics C: Solid State
  Physics}\ }\textbf {\bibinfo {volume} {12}},\ \bibinfo {pages} {L729}
  (\bibinfo {year} {1979})}\BibitemShut {NoStop}%
\end{thebibliography}%


\begin{thebibliography}{40}%
\makeatletter
\providecommand \@ifxundefined [1]{%
 \@ifx{#1\undefined}
}%
\providecommand \@ifnum [1]{%
 \ifnum #1\expandafter \@firstoftwo
 \else \expandafter \@secondoftwo
 \fi
}%
\providecommand \@ifx [1]{%
 \ifx #1\expandafter \@firstoftwo
 \else \expandafter \@secondoftwo
 \fi
}%
\providecommand \natexlab [1]{#1}%
\providecommand \enquote  [1]{``#1''}%
\providecommand \bibnamefont  [1]{#1}%
\providecommand \bibfnamefont [1]{#1}%
\providecommand \citenamefont [1]{#1}%
\providecommand \href@noop [0]{\@secondoftwo}%
\providecommand \href [0]{\begingroup \@sanitize@url \@href}%
\providecommand \@href[1]{\@@startlink{#1}\@@href}%
\providecommand \@@href[1]{\endgroup#1\@@endlink}%
\providecommand \@sanitize@url [0]{\catcode `\\12\catcode `\$12\catcode
  `\&12\catcode `\#12\catcode `\^12\catcode `\_12\catcode `\%12\relax}%
\providecommand \@@startlink[1]{}%
\providecommand \@@endlink[0]{}%
\providecommand \url  [0]{\begingroup\@sanitize@url \@url }%
\providecommand \@url [1]{\endgroup\@href {#1}{\urlprefix }}%
\providecommand \urlprefix  [0]{URL }%
\providecommand \Eprint [0]{\href }%
\providecommand \doibase [0]{https://doi.org/}%
\providecommand \selectlanguage [0]{\@gobble}%
\providecommand \bibinfo  [0]{\@secondoftwo}%
\providecommand \bibfield  [0]{\@secondoftwo}%
\providecommand \translation [1]{[#1]}%
\providecommand \BibitemOpen [0]{}%
\providecommand \bibitemStop [0]{}%
\providecommand \bibitemNoStop [0]{.\EOS\space}%
\providecommand \EOS [0]{\spacefactor3000\relax}%
\providecommand \BibitemShut  [1]{\csname bibitem#1\endcsname}%
\let\auto@bib@innerbib\@empty
\bibitem [{\citenamefont {Sachdev}(2011)}]{Sachdev2011}%
  \BibitemOpen
  \bibfield  {author} {\bibinfo {author} {\bibfnamefont {S.}~\bibnamefont
  {Sachdev}},\ }\href@noop {} {\emph {\bibinfo {title} {Quantum Phase
  Transitions}}}\ (\bibinfo  {publisher} {Cambridge University Press},\
  \bibinfo {address} {Cambridge},\ \bibinfo {year} {2011})\BibitemShut
  {NoStop}%
\bibitem [{\citenamefont {Scalapino}(2012)}]{Scalapino2012}%
  \BibitemOpen
  \bibfield  {author} {\bibinfo {author} {\bibfnamefont {D.~J.}\ \bibnamefont
  {Scalapino}},\ }\href {https://doi.org/10.1103/RevModPhys.84.1383} {\bibfield
   {journal} {\bibinfo  {journal} {Rev. Mod. Phys.}\ }\textbf {\bibinfo
  {volume} {84}},\ \bibinfo {pages} {1383} (\bibinfo {year}
  {2012})}\BibitemShut {NoStop}%
\bibitem [{\citenamefont {Monthoux}\ \emph {et~al.}(2007)\citenamefont
  {Monthoux}, \citenamefont {Pines},\ and\ \citenamefont
  {Lonzarich}}]{Monthoux2007}%
  \BibitemOpen
  \bibfield  {author} {\bibinfo {author} {\bibfnamefont {P.}~\bibnamefont
  {Monthoux}}, \bibinfo {author} {\bibfnamefont {D.}~\bibnamefont {Pines}},\
  and\ \bibinfo {author} {\bibfnamefont {G.~G.}\ \bibnamefont {Lonzarich}},\
  }\href@noop {} {\bibfield  {journal} {\bibinfo  {journal} {Nature}\ }\textbf
  {\bibinfo {volume} {450}},\ \bibinfo {pages} {1177} (\bibinfo {year}
  {2007})}\BibitemShut {NoStop}%
\bibitem [{\citenamefont {Ghiotto}\ \emph {et~al.}(2021)\citenamefont
  {Ghiotto}, \citenamefont {Shih}, \citenamefont {Pereira}, \citenamefont
  {Rhodes}, \citenamefont {Kim}, \citenamefont {Zang}, \citenamefont {Millis},
  \citenamefont {Watanabe}, \citenamefont {Taniguchi}, \citenamefont {Hone},
  \citenamefont {Wang}, \citenamefont {Dean},\ and\ \citenamefont
  {Pasupathy}}]{Ghiotto2021}%
  \BibitemOpen
  \bibfield  {author} {\bibinfo {author} {\bibfnamefont {A.}~\bibnamefont
  {Ghiotto}}, \bibinfo {author} {\bibfnamefont {E.-M.}\ \bibnamefont {Shih}},
  \bibinfo {author} {\bibfnamefont {G.~S. S.~G.}\ \bibnamefont {Pereira}},
  \bibinfo {author} {\bibfnamefont {D.~A.}\ \bibnamefont {Rhodes}}, \bibinfo
  {author} {\bibfnamefont {B.}~\bibnamefont {Kim}}, \bibinfo {author}
  {\bibfnamefont {J.}~\bibnamefont {Zang}}, \bibinfo {author} {\bibfnamefont
  {A.~J.}\ \bibnamefont {Millis}}, \bibinfo {author} {\bibfnamefont
  {K.}~\bibnamefont {Watanabe}}, \bibinfo {author} {\bibfnamefont
  {T.}~\bibnamefont {Taniguchi}}, \bibinfo {author} {\bibfnamefont {J.~C.}\
  \bibnamefont {Hone}}, \bibinfo {author} {\bibfnamefont {L.}~\bibnamefont
  {Wang}}, \bibinfo {author} {\bibfnamefont {C.~R.}\ \bibnamefont {Dean}},\
  and\ \bibinfo {author} {\bibfnamefont {A.~N.}\ \bibnamefont {Pasupathy}},\
  }\href {https://doi.org/10.1038/s41586-021-03815-6} {\bibfield  {journal}
  {\bibinfo  {journal} {Nature}\ }\textbf {\bibinfo {volume} {597}},\ \bibinfo
  {pages} {345} (\bibinfo {year} {2021})}\BibitemShut {NoStop}%
\bibitem [{\citenamefont {Rowley}\ \emph {et~al.}(2014)\citenamefont {Rowley},
  \citenamefont {Spalek}, \citenamefont {Smith}, \citenamefont {Dean},
  \citenamefont {Itoh}, \citenamefont {Scott}, \citenamefont {Lonzarich},\ and\
  \citenamefont {Saxena}}]{Rowley2014}%
  \BibitemOpen
  \bibfield  {author} {\bibinfo {author} {\bibfnamefont {S.~E.}\ \bibnamefont
  {Rowley}}, \bibinfo {author} {\bibfnamefont {L.~J.}\ \bibnamefont {Spalek}},
  \bibinfo {author} {\bibfnamefont {R.~P.}\ \bibnamefont {Smith}}, \bibinfo
  {author} {\bibfnamefont {M.~P.~M.}\ \bibnamefont {Dean}}, \bibinfo {author}
  {\bibfnamefont {M.}~\bibnamefont {Itoh}}, \bibinfo {author} {\bibfnamefont
  {J.~F.}\ \bibnamefont {Scott}}, \bibinfo {author} {\bibfnamefont {G.~G.}\
  \bibnamefont {Lonzarich}},\ and\ \bibinfo {author} {\bibfnamefont {S.~S.}\
  \bibnamefont {Saxena}},\ }\href {https://doi.org/10.1038/nphys2924}
  {\bibfield  {journal} {\bibinfo  {journal} {Nature Physics}\ }\textbf
  {\bibinfo {volume} {10}},\ \bibinfo {pages} {367} (\bibinfo {year}
  {2014})}\BibitemShut {NoStop}%
\bibitem [{\citenamefont {Stockert}\ and\ \citenamefont
  {Steglich}(2011)}]{Stockert2011}%
  \BibitemOpen
  \bibfield  {author} {\bibinfo {author} {\bibfnamefont {O.}~\bibnamefont
  {Stockert}}\ and\ \bibinfo {author} {\bibfnamefont {F.}~\bibnamefont
  {Steglich}},\ }\href
  {https://doi.org/10.1146/annurev-conmatphys-062910-140546} {\bibfield
  {journal} {\bibinfo  {journal} {Annual Review of Condensed Matter Physics}\
  }\textbf {\bibinfo {volume} {2}},\ \bibinfo {pages} {79} (\bibinfo {year}
  {2011})}\BibitemShut {NoStop}%
\bibitem [{\citenamefont {Stewart}(2001)}]{Stewart01}%
  \BibitemOpen
  \bibfield  {author} {\bibinfo {author} {\bibfnamefont {G.~R.}\ \bibnamefont
  {Stewart}},\ }\href@noop {} {\bibfield  {journal} {\bibinfo  {journal} {Rev.
  Mod. Phys.}\ }\textbf {\bibinfo {volume} {73}},\ \bibinfo {pages} {797}
  (\bibinfo {year} {2001})}\BibitemShut {NoStop}%
\bibitem [{\citenamefont {L\"ohneysen}\ \emph {et~al.}(2007)\citenamefont
  {L\"ohneysen}, \citenamefont {Rosch}, \citenamefont {Vojta},\ and\
  \citenamefont {W\"olfle}}]{vonLohneysen07}%
  \BibitemOpen
  \bibfield  {author} {\bibinfo {author} {\bibfnamefont {H.~v.}\ \bibnamefont
  {L\"ohneysen}}, \bibinfo {author} {\bibfnamefont {A.}~\bibnamefont {Rosch}},
  \bibinfo {author} {\bibfnamefont {M.}~\bibnamefont {Vojta}},\ and\ \bibinfo
  {author} {\bibfnamefont {P.}~\bibnamefont {W\"olfle}},\ }\href
  {https://doi.org/10.1103/RevModPhys.79.1015} {\bibfield  {journal} {\bibinfo
  {journal} {Rev. Mod. Phys.}\ }\textbf {\bibinfo {volume} {79}},\ \bibinfo
  {pages} {1015} (\bibinfo {year} {2007})}\BibitemShut {NoStop}%
\bibitem [{\citenamefont {Paschen}\ and\ \citenamefont
  {Si}(2021)}]{Paschen2021}%
  \BibitemOpen
  \bibfield  {author} {\bibinfo {author} {\bibfnamefont {S.}~\bibnamefont
  {Paschen}}\ and\ \bibinfo {author} {\bibfnamefont {Q.}~\bibnamefont {Si}},\
  }\href {https://doi.org/10.1038/s42254-020-00262-6} {\bibfield  {journal}
  {\bibinfo  {journal} {Nature Reviews Physics}\ }\textbf {\bibinfo {volume}
  {3}},\ \bibinfo {pages} {9} (\bibinfo {year} {2021})}\BibitemShut {NoStop}%
\bibitem [{\citenamefont {Millis}(1993)}]{Millis93}%
  \BibitemOpen
  \bibfield  {author} {\bibinfo {author} {\bibfnamefont {A.~J.}\ \bibnamefont
  {Millis}},\ }\href@noop {} {\bibfield  {journal} {\bibinfo  {journal} {Phys.
  Rev. B}\ }\textbf {\bibinfo {volume} {48}},\ \bibinfo {pages} {7183}
  (\bibinfo {year} {1993})}\BibitemShut {NoStop}%
\bibitem [{\citenamefont {Hertz}(1976)}]{Hertz76}%
  \BibitemOpen
  \bibfield  {author} {\bibinfo {author} {\bibfnamefont {J.~A.}\ \bibnamefont
  {Hertz}},\ }\href@noop {} {\bibfield  {journal} {\bibinfo  {journal} {Phys.
  Rev. B}\ }\textbf {\bibinfo {volume} {14}},\ \bibinfo {pages} {1165}
  (\bibinfo {year} {1976})}\BibitemShut {NoStop}%
\bibitem [{\citenamefont {Kadowaki}\ \emph {et~al.}(2006)\citenamefont
  {Kadowaki}, \citenamefont {Tabata}, \citenamefont {Sato}, \citenamefont
  {Aso}, \citenamefont {Raymond},\ and\ \citenamefont
  {Kawarazaki}}]{Kadowaki2006}%
  \BibitemOpen
  \bibfield  {author} {\bibinfo {author} {\bibfnamefont {H.}~\bibnamefont
  {Kadowaki}}, \bibinfo {author} {\bibfnamefont {Y.}~\bibnamefont {Tabata}},
  \bibinfo {author} {\bibfnamefont {M.}~\bibnamefont {Sato}}, \bibinfo {author}
  {\bibfnamefont {N.}~\bibnamefont {Aso}}, \bibinfo {author} {\bibfnamefont
  {S.}~\bibnamefont {Raymond}},\ and\ \bibinfo {author} {\bibfnamefont
  {S.}~\bibnamefont {Kawarazaki}},\ }\href
  {https://doi.org/10.1103/PhysRevLett.96.016401} {\bibfield  {journal}
  {\bibinfo  {journal} {Phys. Rev. Lett.}\ }\textbf {\bibinfo {volume} {96}},\
  \bibinfo {pages} {016401} (\bibinfo {year} {2006})}\BibitemShut {NoStop}%
\bibitem [{\citenamefont {Wang}\ \emph {et~al.}(2011)\citenamefont {Wang},
  \citenamefont {Lawrence}, \citenamefont {Christianson}, \citenamefont
  {Chang}, \citenamefont {Gofryk}, \citenamefont {Bauer}, \citenamefont
  {Ronning}, \citenamefont {Thompson}, \citenamefont {McClellan}, \citenamefont
  {Rodriguez-Rivera},\ and\ \citenamefont {Lynn}}]{Wang2011}%
  \BibitemOpen
  \bibfield  {author} {\bibinfo {author} {\bibfnamefont {C.~H.}\ \bibnamefont
  {Wang}}, \bibinfo {author} {\bibfnamefont {J.~M.}\ \bibnamefont {Lawrence}},
  \bibinfo {author} {\bibfnamefont {A.~D.}\ \bibnamefont {Christianson}},
  \bibinfo {author} {\bibfnamefont {S.}~\bibnamefont {Chang}}, \bibinfo
  {author} {\bibfnamefont {K.}~\bibnamefont {Gofryk}}, \bibinfo {author}
  {\bibfnamefont {E.~D.}\ \bibnamefont {Bauer}}, \bibinfo {author}
  {\bibfnamefont {F.}~\bibnamefont {Ronning}}, \bibinfo {author} {\bibfnamefont
  {J.~D.}\ \bibnamefont {Thompson}}, \bibinfo {author} {\bibfnamefont {K.~J.}\
  \bibnamefont {McClellan}}, \bibinfo {author} {\bibfnamefont {J.~A.}\
  \bibnamefont {Rodriguez-Rivera}},\ and\ \bibinfo {author} {\bibfnamefont
  {J.~W.}\ \bibnamefont {Lynn}},\ }\href
  {https://dx.doi.org/10.1088/1742-6596/273/1/012018} {\bibfield  {journal}
  {\bibinfo  {journal} {Journal of Physics: Conference Series}\ }\textbf
  {\bibinfo {volume} {273}},\ \bibinfo {pages} {012018} (\bibinfo {year}
  {2011})}\BibitemShut {NoStop}%
\bibitem [{\citenamefont {Moriya}\ and\ \citenamefont
  {Takimoto}(1995)}]{Moriya95}%
  \BibitemOpen
  \bibfield  {author} {\bibinfo {author} {\bibfnamefont {T.}~\bibnamefont
  {Moriya}}\ and\ \bibinfo {author} {\bibfnamefont {T.}~\bibnamefont
  {Takimoto}},\ }\href@noop {} {\bibfield  {journal} {\bibinfo  {journal} {J.
  Phys. Soc. Japan}\ }\textbf {\bibinfo {volume} {64}},\ \bibinfo {pages} {960}
  (\bibinfo {year} {1995})},\ \bibinfo {note} {and refs. therein}\BibitemShut
  {NoStop}%
\bibitem [{\citenamefont {Asaba}\ \emph {et~al.}(2021)\citenamefont {Asaba},
  \citenamefont {Lee}, \citenamefont {Seo}, \citenamefont {Avers},
  \citenamefont {Thomas}, \citenamefont {Movshovich}, \citenamefont {Thompson},
  \citenamefont {Rosa}, \citenamefont {Bauer},\ and\ \citenamefont
  {Ronning}}]{Asaba2021}%
  \BibitemOpen
  \bibfield  {author} {\bibinfo {author} {\bibfnamefont {T.}~\bibnamefont
  {Asaba}}, \bibinfo {author} {\bibfnamefont {S.}~\bibnamefont {Lee}}, \bibinfo
  {author} {\bibfnamefont {S.}~\bibnamefont {Seo}}, \bibinfo {author}
  {\bibfnamefont {K.~E.}\ \bibnamefont {Avers}}, \bibinfo {author}
  {\bibfnamefont {S.~M.}\ \bibnamefont {Thomas}}, \bibinfo {author}
  {\bibfnamefont {R.}~\bibnamefont {Movshovich}}, \bibinfo {author}
  {\bibfnamefont {J.~D.}\ \bibnamefont {Thompson}}, \bibinfo {author}
  {\bibfnamefont {P.~F.~S.}\ \bibnamefont {Rosa}}, \bibinfo {author}
  {\bibfnamefont {E.~D.}\ \bibnamefont {Bauer}},\ and\ \bibinfo {author}
  {\bibfnamefont {F.}~\bibnamefont {Ronning}},\ }\href
  {https://doi.org/10.1103/PhysRevB.104.195140} {\bibfield  {journal} {\bibinfo
   {journal} {Phys. Rev. B}\ }\textbf {\bibinfo {volume} {104}},\ \bibinfo
  {pages} {195140} (\bibinfo {year} {2021})}\BibitemShut {NoStop}%
\bibitem [{K.E. Avers et al.()}]{Avers2024}%
  \BibitemOpen
  K.E. Avers et al.,\ \href@noop {} {} (\bibinfo {year} {2024}),\ \bibinfo
  {note} {(unpublished)}\BibitemShut {NoStop}%
\bibitem [{\citenamefont {Matsumoto}\ \emph {et~al.}(2011)\citenamefont
  {Matsumoto}, \citenamefont {Nakatsuji}, \citenamefont {Kuga}, \citenamefont
  {Karaki}, \citenamefont {Horie}, \citenamefont {Shimura}, \citenamefont
  {Sakakibara}, \citenamefont {Nevidomskyy},\ and\ \citenamefont
  {Coleman}}]{Matsumoto2011}%
  \BibitemOpen
  \bibfield  {author} {\bibinfo {author} {\bibfnamefont {Y.}~\bibnamefont
  {Matsumoto}}, \bibinfo {author} {\bibfnamefont {S.}~\bibnamefont
  {Nakatsuji}}, \bibinfo {author} {\bibfnamefont {K.}~\bibnamefont {Kuga}},
  \bibinfo {author} {\bibfnamefont {Y.}~\bibnamefont {Karaki}}, \bibinfo
  {author} {\bibfnamefont {N.}~\bibnamefont {Horie}}, \bibinfo {author}
  {\bibfnamefont {Y.}~\bibnamefont {Shimura}}, \bibinfo {author} {\bibfnamefont
  {T.}~\bibnamefont {Sakakibara}}, \bibinfo {author} {\bibfnamefont {A.~H.}\
  \bibnamefont {Nevidomskyy}},\ and\ \bibinfo {author} {\bibfnamefont
  {P.}~\bibnamefont {Coleman}},\ }\href
  {https://www.science.org/doi/abs/10.1126/science.1197531} {\bibfield
  {journal} {\bibinfo  {journal} {Science}\ }\textbf {\bibinfo {volume}
  {331}},\ \bibinfo {pages} {316} (\bibinfo {year} {2011})}\BibitemShut
  {NoStop}%
\bibitem [{\citenamefont {Wu}\ \emph {et~al.}(2014)\citenamefont {Wu},
  \citenamefont {Kim}, \citenamefont {Park}, \citenamefont {Tsvelik},\ and\
  \citenamefont {Aronson}}]{Wu2014}%
  \BibitemOpen
  \bibfield  {author} {\bibinfo {author} {\bibfnamefont {L.~S.}\ \bibnamefont
  {Wu}}, \bibinfo {author} {\bibfnamefont {M.~S.}\ \bibnamefont {Kim}},
  \bibinfo {author} {\bibfnamefont {K.}~\bibnamefont {Park}}, \bibinfo {author}
  {\bibfnamefont {A.~M.}\ \bibnamefont {Tsvelik}},\ and\ \bibinfo {author}
  {\bibfnamefont {M.~C.}\ \bibnamefont {Aronson}},\ }\href
  {https://www.pnas.org/doi/abs/10.1073/pnas.1413112111} {\bibfield  {journal}
  {\bibinfo  {journal} {Proceedings of the National Academy of Sciences}\
  }\textbf {\bibinfo {volume} {111}},\ \bibinfo {pages} {14088} (\bibinfo
  {year} {2014})}\BibitemShut {NoStop}%
\bibitem [{\citenamefont {Fuhrman}\ \emph {et~al.}(2021)\citenamefont
  {Fuhrman}, \citenamefont {Sidorenko}, \citenamefont {Haenel}, \citenamefont
  {Winkler}, \citenamefont {Prokofiev}, \citenamefont {Rodriguez-Rivera},
  \citenamefont {Qiu}, \citenamefont {Blaha}, \citenamefont {Si}, \citenamefont
  {Broholm},\ and\ \citenamefont {Paschen}}]{Fuhrman2021}%
  \BibitemOpen
  \bibfield  {author} {\bibinfo {author} {\bibfnamefont {W.~T.}\ \bibnamefont
  {Fuhrman}}, \bibinfo {author} {\bibfnamefont {A.}~\bibnamefont {Sidorenko}},
  \bibinfo {author} {\bibfnamefont {J.}~\bibnamefont {Haenel}}, \bibinfo
  {author} {\bibfnamefont {H.}~\bibnamefont {Winkler}}, \bibinfo {author}
  {\bibfnamefont {A.}~\bibnamefont {Prokofiev}}, \bibinfo {author}
  {\bibfnamefont {J.~A.}\ \bibnamefont {Rodriguez-Rivera}}, \bibinfo {author}
  {\bibfnamefont {Y.}~\bibnamefont {Qiu}}, \bibinfo {author} {\bibfnamefont
  {P.}~\bibnamefont {Blaha}}, \bibinfo {author} {\bibfnamefont
  {Q.}~\bibnamefont {Si}}, \bibinfo {author} {\bibfnamefont {C.~L.}\
  \bibnamefont {Broholm}},\ and\ \bibinfo {author} {\bibfnamefont
  {S.}~\bibnamefont {Paschen}},\ }\href
  {https://doi.org/10.1126/sciadv.abf9134} {\bibfield  {journal} {\bibinfo
  {journal} {Science Advances}\ }\textbf {\bibinfo {volume} {7}},\ \bibinfo
  {pages} {eabf9134} (\bibinfo {year} {2021})}\BibitemShut {NoStop}%
\bibitem [{\citenamefont {Wu}\ \emph {et~al.}(2019)\citenamefont {Wu},
  \citenamefont {Nikitin}, \citenamefont {Wang}, \citenamefont {Zhu},
  \citenamefont {Batista}, \citenamefont {Tsvelik}, \citenamefont {Samarakoon},
  \citenamefont {Tennant}, \citenamefont {Brando}, \citenamefont {Vasylechko},
  \citenamefont {Frontzek}, \citenamefont {Savici}, \citenamefont {Sala},
  \citenamefont {Ehlers}, \citenamefont {Christianson}, \citenamefont
  {Lumsden},\ and\ \citenamefont {Podlesnyak}}]{Wu2019}%
  \BibitemOpen
  \bibfield  {author} {\bibinfo {author} {\bibfnamefont {L.~S.}\ \bibnamefont
  {Wu}}, \bibinfo {author} {\bibfnamefont {S.~E.}\ \bibnamefont {Nikitin}},
  \bibinfo {author} {\bibfnamefont {Z.}~\bibnamefont {Wang}}, \bibinfo {author}
  {\bibfnamefont {W.}~\bibnamefont {Zhu}}, \bibinfo {author} {\bibfnamefont
  {C.~D.}\ \bibnamefont {Batista}}, \bibinfo {author} {\bibfnamefont {A.~M.}\
  \bibnamefont {Tsvelik}}, \bibinfo {author} {\bibfnamefont {A.~M.}\
  \bibnamefont {Samarakoon}}, \bibinfo {author} {\bibfnamefont {D.~A.}\
  \bibnamefont {Tennant}}, \bibinfo {author} {\bibfnamefont {M.}~\bibnamefont
  {Brando}}, \bibinfo {author} {\bibfnamefont {L.}~\bibnamefont {Vasylechko}},
  \bibinfo {author} {\bibfnamefont {M.}~\bibnamefont {Frontzek}}, \bibinfo
  {author} {\bibfnamefont {A.~T.}\ \bibnamefont {Savici}}, \bibinfo {author}
  {\bibfnamefont {G.}~\bibnamefont {Sala}}, \bibinfo {author} {\bibfnamefont
  {G.}~\bibnamefont {Ehlers}}, \bibinfo {author} {\bibfnamefont {A.~D.}\
  \bibnamefont {Christianson}}, \bibinfo {author} {\bibfnamefont {M.~D.}\
  \bibnamefont {Lumsden}},\ and\ \bibinfo {author} {\bibfnamefont
  {A.}~\bibnamefont {Podlesnyak}},\ }\href
  {https://doi.org/10.1038/s41467-019-08485-7} {\bibfield  {journal} {\bibinfo
  {journal} {Nature Communications}\ }\textbf {\bibinfo {volume} {10}},\
  \bibinfo {pages} {698} (\bibinfo {year} {2019})}\BibitemShut {NoStop}%
\bibitem [{SM()}]{SM}%
  \BibitemOpen
  \href@noop {} {}\bibinfo {note} {See Supplemental Material at ... for details
  of: a schematic phase diagram for Renormalization Group flows of an
  antiferromagnetic near a QCP in a uniform magnetic field, the magnetic
  Gr\"uneisen parameter as a function of magnetic field and temperature,
  scaling analysis for H$\parallel$[201] and H$\parallel$b, scaling curve for
  specific heat from a Maxwell relation, schematic H-T-P (y) phase diagram for
  YbFe$_5$P$_3$ near an AFMQCP, a detailed analysis of Extended Mean Field vs.
  Gaussian behavior near a QCP, other possibilities for the QCP in
  YbFe$_5$P$_3$, and a scaling analysis of CeRu$_6$Sn$_4$.}\BibitemShut {Stop}%
\bibitem [{\citenamefont {Ma}(1985)}]{Ma1985}%
  \BibitemOpen
  \bibfield  {author} {\bibinfo {author} {\bibfnamefont {S.-K.}\ \bibnamefont
  {Ma}},\ }\href {https://doi.org/10.1142/0073} {\emph {\bibinfo {title}
  {Statistical Mechanics}}}\ (\bibinfo  {publisher} {World Scientific},\
  \bibinfo {year} {1985})\ \Eprint
  {https://arxiv.org/abs/https://www.worldscientific.com/doi/pdf/10.1142/0073}
  {https://www.worldscientific.com/doi/pdf/10.1142/0073} \BibitemShut {NoStop}%
\bibitem [{\citenamefont {Ferreira}\ \emph {et~al.}(1991)\citenamefont
  {Ferreira}, \citenamefont {King},\ and\ \citenamefont
  {Jaccarino}}]{Ferreira1991}%
  \BibitemOpen
  \bibfield  {author} {\bibinfo {author} {\bibfnamefont {I.~B.}\ \bibnamefont
  {Ferreira}}, \bibinfo {author} {\bibfnamefont {A.~R.}\ \bibnamefont {King}},\
  and\ \bibinfo {author} {\bibfnamefont {V.}~\bibnamefont {Jaccarino}},\ }\href
  {https://doi.org/10.1103/PhysRevB.43.10797} {\bibfield  {journal} {\bibinfo
  {journal} {Phys. Rev. B}\ }\textbf {\bibinfo {volume} {43}},\ \bibinfo
  {pages} {10797} (\bibinfo {year} {1991})}\BibitemShut {NoStop}%
\bibitem [{\citenamefont {Fishman}\ and\ \citenamefont
  {Aharony}(1979)}]{Fishman1979}%
  \BibitemOpen
  \bibfield  {author} {\bibinfo {author} {\bibfnamefont {S.}~\bibnamefont
  {Fishman}}\ and\ \bibinfo {author} {\bibfnamefont {A.}~\bibnamefont
  {Aharony}},\ }\href {https://dx.doi.org/10.1088/0022-3719/12/18/006}
  {\bibfield  {journal} {\bibinfo  {journal} {Journal of Physics C: Solid State
  Physics}\ }\textbf {\bibinfo {volume} {12}},\ \bibinfo {pages} {L729}
  (\bibinfo {year} {1979})}\BibitemShut {NoStop}%
\bibitem [{\citenamefont {Garst}\ and\ \citenamefont
  {Rosch}(2005)}]{Garst2005}%
  \BibitemOpen
  \bibfield  {author} {\bibinfo {author} {\bibfnamefont {M.}~\bibnamefont
  {Garst}}\ and\ \bibinfo {author} {\bibfnamefont {A.}~\bibnamefont {Rosch}},\
  }\href {https://doi.org/10.1103/PhysRevB.72.205129} {\bibfield  {journal}
  {\bibinfo  {journal} {Phys. Rev. B}\ }\textbf {\bibinfo {volume} {72}},\
  \bibinfo {pages} {205129} (\bibinfo {year} {2005})}\BibitemShut {NoStop}%
\bibitem [{\citenamefont {Tokiwa}\ \emph {et~al.}(2009)\citenamefont {Tokiwa},
  \citenamefont {Radu}, \citenamefont {Geibel}, \citenamefont {Steglich},\ and\
  \citenamefont {Gegenwart}}]{Tokiwa09}%
  \BibitemOpen
  \bibfield  {author} {\bibinfo {author} {\bibfnamefont {Y.}~\bibnamefont
  {Tokiwa}}, \bibinfo {author} {\bibfnamefont {T.}~\bibnamefont {Radu}},
  \bibinfo {author} {\bibfnamefont {C.}~\bibnamefont {Geibel}}, \bibinfo
  {author} {\bibfnamefont {F.}~\bibnamefont {Steglich}},\ and\ \bibinfo
  {author} {\bibfnamefont {P.}~\bibnamefont {Gegenwart}},\ }\href
  {https://doi.org/10.1103/PhysRevLett.102.066401} {\bibfield  {journal}
  {\bibinfo  {journal} {Phys. Rev. Lett.}\ }\textbf {\bibinfo {volume} {102}},\
  \bibinfo {pages} {066401} (\bibinfo {year} {2009})}\BibitemShut {NoStop}%
\bibitem [{\citenamefont {Danu}\ \emph {et~al.}(2022)\citenamefont {Danu},
  \citenamefont {Vojta}, \citenamefont {Grover},\ and\ \citenamefont
  {Assaad}}]{Danu2022}%
  \BibitemOpen
  \bibfield  {author} {\bibinfo {author} {\bibfnamefont {B.}~\bibnamefont
  {Danu}}, \bibinfo {author} {\bibfnamefont {M.}~\bibnamefont {Vojta}},
  \bibinfo {author} {\bibfnamefont {T.}~\bibnamefont {Grover}},\ and\ \bibinfo
  {author} {\bibfnamefont {F.~F.}\ \bibnamefont {Assaad}},\ }\href
  {https://doi.org/10.1103/PhysRevB.106.L161103} {\bibfield  {journal}
  {\bibinfo  {journal} {Phys. Rev. B}\ }\textbf {\bibinfo {volume} {106}},\
  \bibinfo {pages} {L161103} (\bibinfo {year} {2022})}\BibitemShut {NoStop}%
\bibitem [{\citenamefont {Giamarchi}(2003)}]{Giamarchi2003}%
  \BibitemOpen
  \bibfield  {author} {\bibinfo {author} {\bibfnamefont {T.}~\bibnamefont
  {Giamarchi}},\ }\href
  {https://doi.org/10.1093/acprof:oso/9780198525004.001.0001} {\emph {\bibinfo
  {title} {{Quantum Physics in One Dimension}}}}\ (\bibinfo  {publisher}
  {Oxford University Press},\ \bibinfo {year} {2003})\BibitemShut {NoStop}%
\bibitem [{\citenamefont {Birgeneau}\ and\ \citenamefont
  {Shirane}(1978)}]{Birgenau1978}%
  \BibitemOpen
  \bibfield  {author} {\bibinfo {author} {\bibfnamefont {R.~J.}\ \bibnamefont
  {Birgeneau}}\ and\ \bibinfo {author} {\bibfnamefont {G.}~\bibnamefont
  {Shirane}},\ }\href {https://doi.org/10.1063/1.2994868} {\bibfield  {journal}
  {\bibinfo  {journal} {Physics Today}\ }\textbf {\bibinfo {volume} {31}},\
  \bibinfo {pages} {32} (\bibinfo {year} {1978})}\BibitemShut {NoStop}%
\bibitem [{\citenamefont {Schröder}\ \emph {et~al.}(2000)\citenamefont
  {Schröder}, \citenamefont {Aeppli}, \citenamefont {Coldea}, \citenamefont
  {Adams}, \citenamefont {Stockert}, \citenamefont {Löhneysen}, \citenamefont
  {Bucher}, \citenamefont {Ramazashvili},\ and\ \citenamefont
  {Coleman}}]{Schroder2000}%
  \BibitemOpen
  \bibfield  {author} {\bibinfo {author} {\bibfnamefont {A.}~\bibnamefont
  {Schröder}}, \bibinfo {author} {\bibfnamefont {G.}~\bibnamefont {Aeppli}},
  \bibinfo {author} {\bibfnamefont {R.}~\bibnamefont {Coldea}}, \bibinfo
  {author} {\bibfnamefont {M.}~\bibnamefont {Adams}}, \bibinfo {author}
  {\bibfnamefont {O.}~\bibnamefont {Stockert}}, \bibinfo {author}
  {\bibfnamefont {H.~v.}\ \bibnamefont {Löhneysen}}, \bibinfo {author}
  {\bibfnamefont {E.}~\bibnamefont {Bucher}}, \bibinfo {author} {\bibfnamefont
  {R.}~\bibnamefont {Ramazashvili}},\ and\ \bibinfo {author} {\bibfnamefont
  {P.}~\bibnamefont {Coleman}},\ }\href {https://doi.org/10.1038/35030039}
  {\bibfield  {journal} {\bibinfo  {journal} {Nature}\ }\textbf {\bibinfo
  {volume} {407}},\ \bibinfo {pages} {351} (\bibinfo {year}
  {2000})}\BibitemShut {NoStop}%
\bibitem [{\citenamefont {Custers}\ \emph {et~al.}(2003)\citenamefont
  {Custers}, \citenamefont {Gegenwart}, \citenamefont {Wilhelm}, \citenamefont
  {Neumaier}, \citenamefont {Tokiwa}, \citenamefont {Trovarelli}, \citenamefont
  {Geibel}, \citenamefont {Steglich}, \citenamefont {P{\'e}pin},\ and\
  \citenamefont {Coleman}}]{Custers03}%
  \BibitemOpen
  \bibfield  {author} {\bibinfo {author} {\bibfnamefont {J.}~\bibnamefont
  {Custers}}, \bibinfo {author} {\bibfnamefont {P.}~\bibnamefont {Gegenwart}},
  \bibinfo {author} {\bibfnamefont {H.}~\bibnamefont {Wilhelm}}, \bibinfo
  {author} {\bibfnamefont {K.}~\bibnamefont {Neumaier}}, \bibinfo {author}
  {\bibfnamefont {Y.}~\bibnamefont {Tokiwa}}, \bibinfo {author} {\bibfnamefont
  {O.}~\bibnamefont {Trovarelli}}, \bibinfo {author} {\bibfnamefont
  {C.}~\bibnamefont {Geibel}}, \bibinfo {author} {\bibfnamefont
  {F.}~\bibnamefont {Steglich}}, \bibinfo {author} {\bibfnamefont
  {C.}~\bibnamefont {P{\'e}pin}},\ and\ \bibinfo {author} {\bibfnamefont
  {P.}~\bibnamefont {Coleman}},\ }\href@noop {} {\bibfield  {journal} {\bibinfo
   {journal} {Nature}\ }\textbf {\bibinfo {volume} {424}},\ \bibinfo {pages}
  {524} (\bibinfo {year} {2003})}\BibitemShut {NoStop}%
\bibitem [{\citenamefont {Gannon}\ \emph {et~al.}(2018)\citenamefont {Gannon},
  \citenamefont {Wu}, \citenamefont {Zaliznyak}, \citenamefont {Xu},
  \citenamefont {Tsvelik}, \citenamefont {Qiu}, \citenamefont
  {Rodriguez-Rivera},\ and\ \citenamefont {Aronson}}]{Gannon2018}%
  \BibitemOpen
  \bibfield  {author} {\bibinfo {author} {\bibfnamefont {W.~J.}\ \bibnamefont
  {Gannon}}, \bibinfo {author} {\bibfnamefont {L.~S.}\ \bibnamefont {Wu}},
  \bibinfo {author} {\bibfnamefont {I.~A.}\ \bibnamefont {Zaliznyak}}, \bibinfo
  {author} {\bibfnamefont {W.~H.}\ \bibnamefont {Xu}}, \bibinfo {author}
  {\bibfnamefont {A.~M.}\ \bibnamefont {Tsvelik}}, \bibinfo {author}
  {\bibfnamefont {Y.}~\bibnamefont {Qiu}}, \bibinfo {author} {\bibfnamefont
  {J.~A.}\ \bibnamefont {Rodriguez-Rivera}},\ and\ \bibinfo {author}
  {\bibfnamefont {M.~C.}\ \bibnamefont {Aronson}},\ }\href
  {https://www.pnas.org/doi/abs/10.1073/pnas.1721493115} {\bibfield  {journal}
  {\bibinfo  {journal} {Proceedings of the National Academy of Sciences}\
  }\textbf {\bibinfo {volume} {115}},\ \bibinfo {pages} {6995} (\bibinfo {year}
  {2018})}\BibitemShut {NoStop}%
\bibitem [{\citenamefont {Continentino}(2017)}]{Continentino2017}%
  \BibitemOpen
  \bibfield  {author} {\bibinfo {author} {\bibfnamefont {M.}~\bibnamefont
  {Continentino}},\ }\href@noop {} {\emph {\bibinfo {title} {Quantum Scaling in
  Many-Body Systems: An Approach to Quantum Phase Transitions}}},\ \bibinfo
  {edition} {2nd}\ ed.\ (\bibinfo  {publisher} {Cambridge University Press},\
  \bibinfo {year} {2017})\BibitemShut {NoStop}%
\bibitem [{\citenamefont {P\'epin}(2007)}]{Pepin2007}%
  \BibitemOpen
  \bibfield  {author} {\bibinfo {author} {\bibfnamefont {C.}~\bibnamefont
  {P\'epin}},\ }\href {https://doi.org/10.1103/PhysRevLett.98.206401}
  {\bibfield  {journal} {\bibinfo  {journal} {Phys. Rev. Lett.}\ }\textbf
  {\bibinfo {volume} {98}},\ \bibinfo {pages} {206401} (\bibinfo {year}
  {2007})}\BibitemShut {NoStop}%
\bibitem [{\citenamefont {Vojta}(2010)}]{Vojta2010}%
  \BibitemOpen
  \bibfield  {author} {\bibinfo {author} {\bibfnamefont {M.}~\bibnamefont
  {Vojta}},\ }\href {https://doi.org/10.1007/s10909-010-0206-3} {\bibfield
  {journal} {\bibinfo  {journal} {Journal of Low Temperature Physics}\ }\textbf
  {\bibinfo {volume} {161}},\ \bibinfo {pages} {203} (\bibinfo {year}
  {2010})}\BibitemShut {NoStop}%
\bibitem [{\citenamefont {de' Medici}\ \emph {et~al.}(2005)\citenamefont {de'
  Medici}, \citenamefont {Georges}, \citenamefont {Kotliar},\ and\
  \citenamefont {Biermann}}]{deMedici2005}%
  \BibitemOpen
  \bibfield  {author} {\bibinfo {author} {\bibfnamefont {L.}~\bibnamefont {de'
  Medici}}, \bibinfo {author} {\bibfnamefont {A.}~\bibnamefont {Georges}},
  \bibinfo {author} {\bibfnamefont {G.}~\bibnamefont {Kotliar}},\ and\ \bibinfo
  {author} {\bibfnamefont {S.}~\bibnamefont {Biermann}},\ }\href
  {https://doi.org/10.1103/PhysRevLett.95.066402} {\bibfield  {journal}
  {\bibinfo  {journal} {Phys. Rev. Lett.}\ }\textbf {\bibinfo {volume} {95}},\
  \bibinfo {pages} {066402} (\bibinfo {year} {2005})}\BibitemShut {NoStop}%
\bibitem [{\citenamefont {Poudel}\ \emph {et~al.}(2019)\citenamefont {Poudel},
  \citenamefont {Lawrence}, \citenamefont {Wu}, \citenamefont {Ehlers},
  \citenamefont {Qiu}, \citenamefont {May}, \citenamefont {Ronning},
  \citenamefont {Lumsden}, \citenamefont {Mandrus},\ and\ \citenamefont
  {Christianson}}]{Poudel2019}%
  \BibitemOpen
  \bibfield  {author} {\bibinfo {author} {\bibfnamefont {L.}~\bibnamefont
  {Poudel}}, \bibinfo {author} {\bibfnamefont {J.~M.}\ \bibnamefont
  {Lawrence}}, \bibinfo {author} {\bibfnamefont {L.~S.}\ \bibnamefont {Wu}},
  \bibinfo {author} {\bibfnamefont {G.}~\bibnamefont {Ehlers}}, \bibinfo
  {author} {\bibfnamefont {Y.}~\bibnamefont {Qiu}}, \bibinfo {author}
  {\bibfnamefont {A.~F.}\ \bibnamefont {May}}, \bibinfo {author} {\bibfnamefont
  {F.}~\bibnamefont {Ronning}}, \bibinfo {author} {\bibfnamefont {M.~D.}\
  \bibnamefont {Lumsden}}, \bibinfo {author} {\bibfnamefont {D.}~\bibnamefont
  {Mandrus}},\ and\ \bibinfo {author} {\bibfnamefont {A.~D.}\ \bibnamefont
  {Christianson}},\ }\href {https://doi.org/10.1038/s41535-019-0191-y}
  {\bibfield  {journal} {\bibinfo  {journal} {npj Quantum Materials}\ }\textbf
  {\bibinfo {volume} {4}},\ \bibinfo {pages} {52} (\bibinfo {year}
  {2019})}\BibitemShut {NoStop}%
\bibitem [{\citenamefont {Continentino}(2001)}]{Continentino2001}%
  \BibitemOpen
  \bibfield  {author} {\bibinfo {author} {\bibfnamefont {M.~A.}\ \bibnamefont
  {Continentino}},\ }\href {https://doi.org/10.1142/4498} {\emph {\bibinfo
  {title} {Quantum Scaling in Many-Body Systems}}}\ (\bibinfo  {publisher}
  {World Scientific},\ \bibinfo {year} {2001})\ \BibitemShut {NoStop}%
\bibitem [{\citenamefont {Ramazashvili}(1999)}]{Ramazashvili1999}%
  \BibitemOpen
  \bibfield  {author} {\bibinfo {author} {\bibfnamefont {R.}~\bibnamefont
  {Ramazashvili}},\ }\href {https://doi.org/10.1103/PhysRevB.60.7314}
  {\bibfield  {journal} {\bibinfo  {journal} {Phys. Rev. B}\ }\textbf {\bibinfo
  {volume} {60}},\ \bibinfo {pages} {7314} (\bibinfo {year}
  {1999})}\BibitemShut {NoStop}%
\bibitem [{\citenamefont {Continentino}(2004)}]{Continentino2004}%
  \BibitemOpen
  \bibfield  {author} {\bibinfo {author} {\bibfnamefont {M.~A.}\ \bibnamefont
  {Continentino}},\ }\href
  {https://doi.org/https://doi.org/10.1016/j.jmmm.2003.11.097} {\bibfield
  {journal} {\bibinfo  {journal} {J. Magn. \& Magn. Mater.}\
  }\textbf {\bibinfo {volume} {272-276}},\ \bibinfo {pages} {231} (\bibinfo
  {year} {2004})}\BibitemShut {NoStop}%
\end{thebibliography}%

\clearpage
\appendix

\bibliographystyle{apsrev4-2}

\setcounter{figure}{0}
\renewcommand{\figurename}{Fig.}
\renewcommand{\thefigure}{S\arabic{figure}}

\textbf{Supplemental Material:  Quantum Critical Scaling in Quasi-One-Dimensional YbFe$_5$P$_3$}

\section{Schematic phase diagram of an antiferromagnet  close to an AFQCP in a uniform magnetic field}

\begin{figure*}[ht]
\centerline{\includegraphics[width=4in]{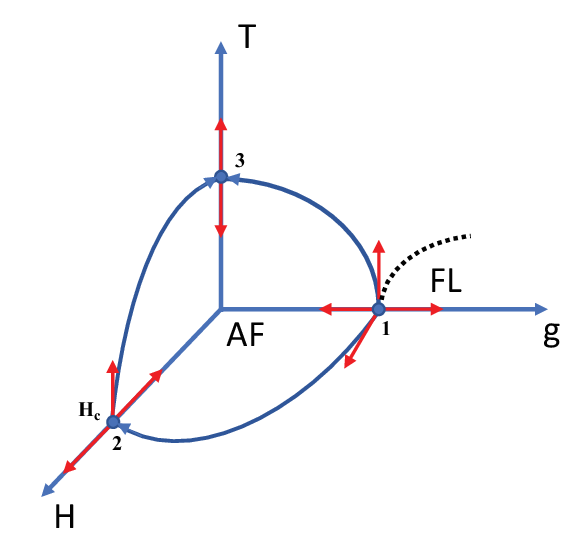}}
\caption{Schematic phase diagram for an antiferromagnet close to a quantum critical point. Fixed point 1 is the usual AFQCP. Fixed point 3 controls the thermal phase transitions in the absence of a magnetic field since temperature is also relevant. The QCP at fixed point 2 governs the zero-temperature transitions in the presence of a magnetic field.}
\label{RG_phase}
\end{figure*}

In Fig. \ref{RG_phase} we show a schematic phase diagram of the fixed points and renormalization group (RG) flows of an antiferromagnet in an external uniform magnetic field. (For simplicity we do not consider any tricritical behavior.) The arrows illustrate the flow of the RG equations.  The fully unstable fixed point 1 is associated with an antiferromagnetic quantum critical point (AFQCP), driven by pressure or alloying, that separates a magnetic phase from a disordered magnetic phase with no long range order. Close to this fixed point, temperature is a relevant field in the RG sense that scales as, $T^\prime=b^z T$, where $z$ is the usual dynamic exponent and $b$ the length scale transformation of the RG procedure. This implies that all finite temperature, zero field transitions are governed by the thermal fixed point 3, and consequently are in a different universality class of the quantum phase transitions. The flow away from $T=0$ on the temperature axis implies that temperature is relevant at the QCP. In the paramagnetic side of the phase diagram, the crossover line in the $T-g$ plane marks the onset of the Fermi liquid regime is governed by the equation:  
\begin{equation}
    T_{coh} = g^{\nu z}  	
\label{Tcohv2}
\end{equation}	 
(not shown). For $d+z > 4$, there is additional line in this paramagnetic side, with exponent $\psi =z/(z+d-2)$, as described by Millis,\cite{Millis93} above which the critical exponents are renormalized by thermal fluctuations (not shown). At $T=0$, in the $H-g$ plane the arrow implies that the magnetic field is a relevant perturbation at the AFQCP. The $T =0$ phase transitions in magnetic field are governed by fixed point 2 and are in a different universality class than that of the zero field AFQCP. The crossover exponent in the $H-g$ plane is $\phi = \nu z_h$, where $z_h$ (or $\phi$) is a new critical exponent. For $z_h = z$, we recover $T/H$ scaling.

\section{Magnetic Gr\"{u}neisen Parameter $\Gamma_{mag}(H,T)$}
\begin{figure*}[ht]
\centerline{\includegraphics[width=4.5in]{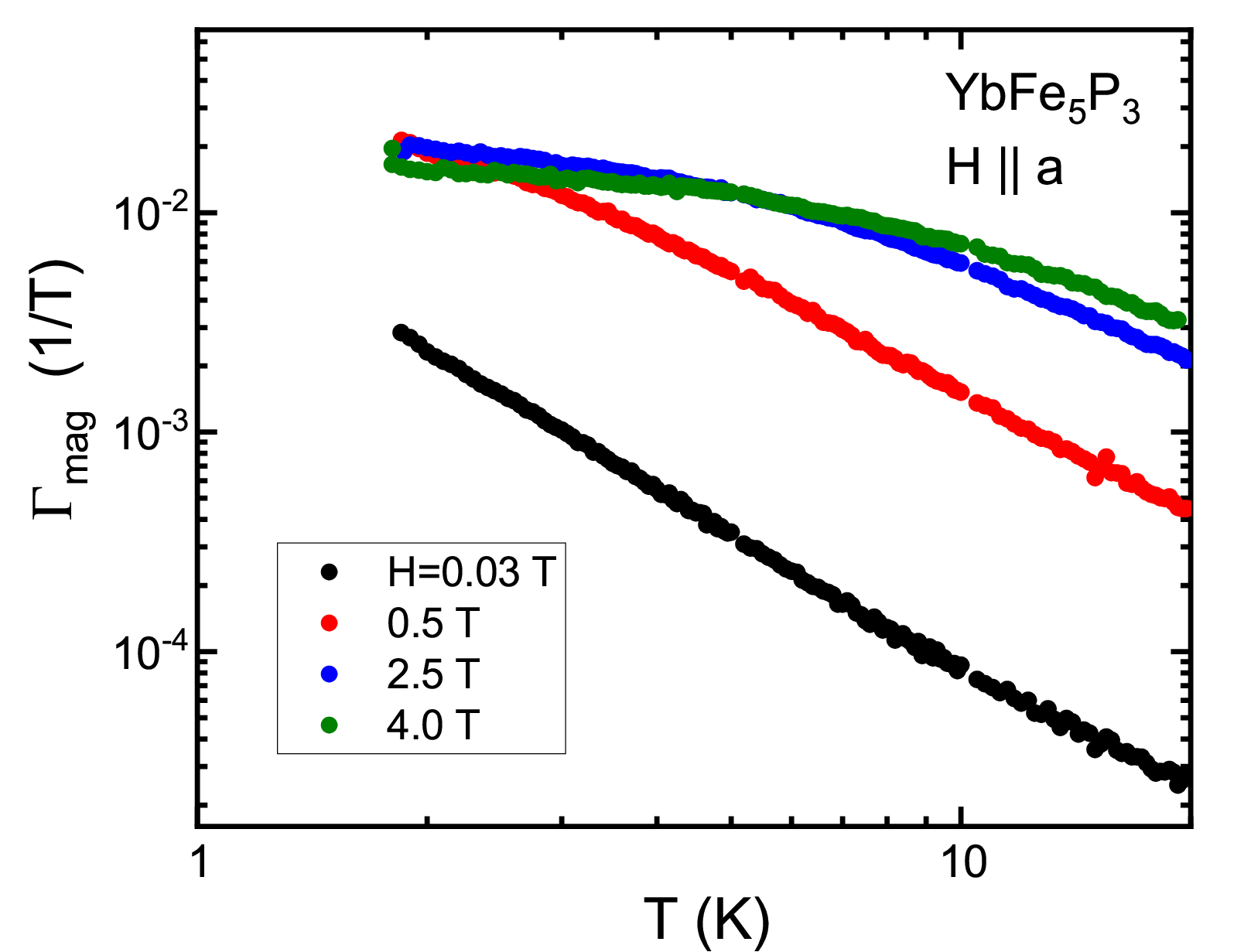}}
\caption{Magnetic Gr\"{u}neisen parameter $\Gamma_{mag}$ vs T in various magnetic fields with H $\parallel a$.}
\label{Gamma_mag_H}
\end{figure*}

In Fig. \ref{Gamma_mag_H}, the magnetic Gr\"{u}neisen parameter $\Gamma_{mag}(H, T)$ of \ybfep{} is displayed for H $\parallel a$ on a semi-log plot for fields up to 4 T.  $\Gamma_{mag}$(H,T) tends to saturate with the application of larger magnetic fields. 

Considering the leading temperature dependence of $\Gamma_{mag}$, we start with the explicit form of the magnetization: 
\begin{equation}
    M =   H^{\frac{2-\alpha -\phi}{\phi}} F_1\left[\frac{T}{H^{\frac{\theta}{\phi}}}\right],  
\mathrm{with} \,\,\, F_1(x) = c(x^2+a^2)^{-\gamma/2}, 
\label{MagnSimpleSM}
\end{equation}	
where $x=T/H^{\theta/\phi}$. Taking a derivative, we have:
\begin{equation}
   \frac{dM}{dT} = \frac{dM}{dx} \frac{dx}{dT} =  - c \gamma x H^{\frac{2-\alpha -\theta -2\phi}{\phi}} (x^2+a^2)^{-\gamma/2-1}.
\label{dMdTSimple}
\end{equation}	
\begin{equation}
  - \frac{1}{T}\frac{dM}{dT} =  - c \gamma H^{-\theta/\phi} H^{\frac{2-\alpha -\theta -2\phi}{\phi}} (x^2+a^2)^{-\gamma/2-1}.
\label{dMdTSimple2}
\end{equation}	
With
\begin{equation}
    \frac{C}{T} = H^{\frac{2-\alpha -2\theta}{\phi}} F_4\left[\frac{T}{H^{\frac{\theta}{\phi}}}\right], 	
\label{CovTSM}
\end{equation}	
we have
\begin{equation}
 \Gamma_{mag} =   \frac{-\frac{1}{T}\frac{dM}{dT}}{\frac{C}{T}} =  - c \gamma H^{\theta/\phi -1}(x^2+a^2)^{-\gamma/2-1}.
\label{Gamma_mag}
\end{equation}	
For $x\gg a$ (in the quantum critical regime), $\gamma=2/3$, and $C/T \sim -ln(T/T_K)$ for $H\rightarrow0$, we have
\begin{equation}
 \Gamma_{mag} =   \frac{(T^2)^{-4/3}}{-ln(T/T_K)}.
\label{Gamma_mag_simple}
\end{equation}	
Using $T_K=$8 K, which is in agreement with the $C_{4f}/T$ data, the expression above reproduces the 1/$T^2$ dependence of $\Gamma_{mag}$.  

\section{Scaling Results for $H\parallel$[201]}
\begin{figure*}[ht]
\centerline{\includegraphics[width=5in]{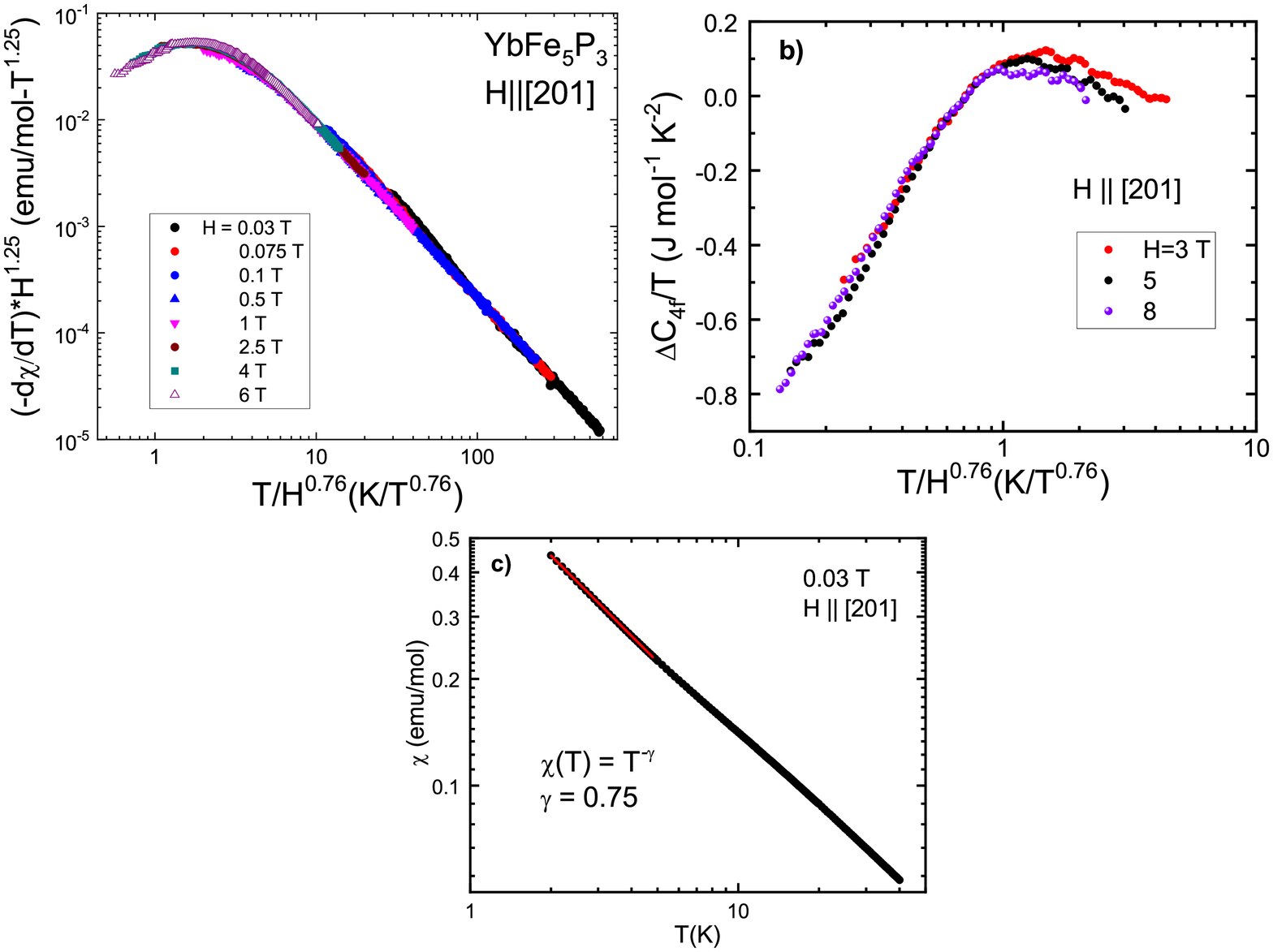}}
\caption{a) Scaling collapse of $d\chi/dT$ vs $T/H^{0.76}$ of \ybfep{} in various applied magnetic fields up to 6 T with H $\parallel [201]$. b)  Scaling collapse of $4f$ contribution to the specific heat, $\Delta C_{4f}/T$ vs $T/H^{0.76}$, in various applied magnetic fields with H $\parallel [201]$. c) Powerlaw fit to $\chi(T)=T^{-\gamma}$ yielding an exponent $\gamma=0.75$, as discussed in the text.   }
\label{scaling201}
\end{figure*}
In Fig. \ref{scaling201}, we present the scaling analysis of the $d\chi(H, T)/dT$ and $C(H,T)/T$ data of \ybfep{} for H $\parallel[201]$. Following the same analysis used for H $\parallel a$ described in the main text, we obtain identical results for H $\parallel[201]$:   
\begin{align}
        & \nu z/\phi=0.76,\nonumber\\ 
   & 2- \frac{\nu d}{\phi}=1.25, \rightarrow  \frac{\nu d}{\phi}=0.75,\nonumber\\ 
    &\frac{\nu(z-d)}{\phi}=0.01
    \label{exponents201}
\end{align}
 The scaling analysis yields $d=z$, as for  H $\parallel a$.  At the QCP, the magnetic susceptibility has the form (in the limit of small fields):
\begin{equation}
    \chi = T^{-\gamma}, 
\label{chi2_201}
\end{equation}	
where $\gamma= \frac{2\phi   -(2-\alpha)}{\theta}$. The specific heat scaling of $\Delta C_{4f}/T$ vs $T/H^{0.76}$ is shown in Fig. \ref{scaling201}c, where $\Delta C_{4f}/T$ = $H^0 F_4(T/H^{{\theta}/\phi})$, similar to the results for H $\parallel a$.
These values imply that the exponent of the uniform susceptibility has the value:
\begin{equation}
     \gamma= \frac{2\phi}{\nu z} -  \frac{(d+z)}{z} =2/3.
\label{gamma_201}
\end{equation}	
In a small magnetic field H=0.03 T, $\chi(T)$ exhibits a powerlaw temperature dependence as shown in Fig. \ref{scaling201}c, albeit over a small temperature range 1.8 K $<$ T $<$ 4.5 K and yields the observed value $\gamma = 0.75$, close to the expected value 2/3.

\section{Scaling Results for $H\parallel$b}
\begin{figure*}[ht]
\centerline{\includegraphics[width=4.5in]{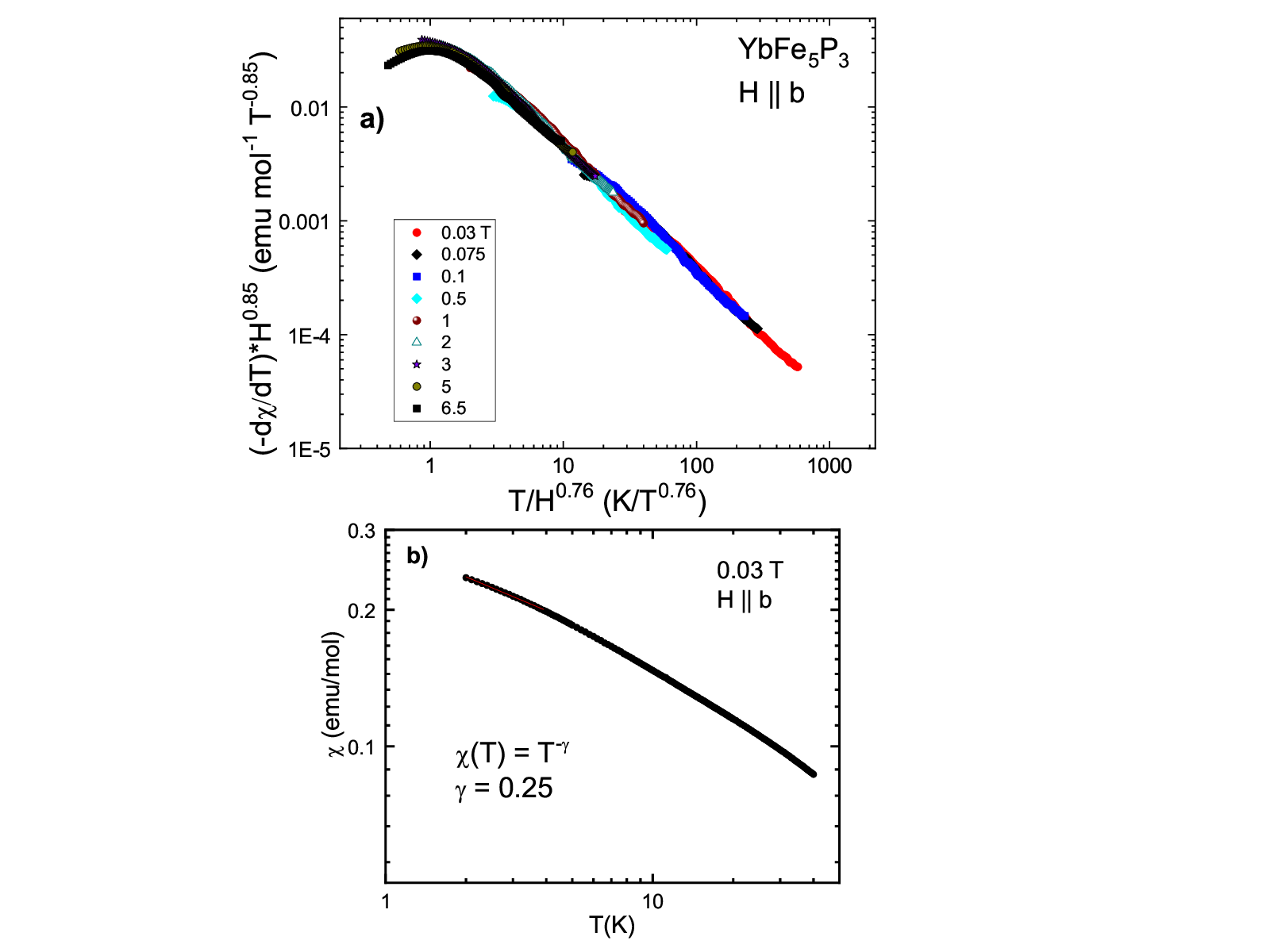}}
\caption{a) Scaling collapse of $d\chi/dT$ vs $T/H^{0.85}$ of \ybfep{} in various applied magnetic fields up to 6.5 T with H $\parallel b$. b)   Powerlaw fit to $\chi(T)=T^{-\gamma}$ yielding an exponent $\gamma=0.25$, as discussed in the text.   }
\label{scalingb}
\end{figure*}
In Fig. \ref{scalingb}, we present the scaling analysis of the $d\chi(H, T)/dT$ and $C(H,T)/T$ data of \ybfep{} for H $\parallel b$. Following the same analysis used for H $\parallel a$ described in the main text, we obtain similar results for H $\parallel b$:   
\begin{align}
        & \nu z/\phi=0.76,\nonumber\\ 
   & 2- \frac{\nu d}{\phi}=0.85, \rightarrow  \frac{\nu d}{\phi}=1.15,\nonumber\\ 
    \label{exponents_b}
\end{align}
 The scaling analysis yields $z=0.66 d$.  For both the d=1 and d=2 cases, the system remains in the hyperscaling regime. At the QCP, the magnetic susceptibility has the form:
\begin{equation}
    \chi = T^{-\gamma}, 
\label{chi2SM}
\end{equation}	
where $\gamma= \frac{2\phi   -(2-\alpha)}{\theta}$.
These values imply that the exponent of the uniform susceptibility has the value:
\begin{equation}
     \gamma= \frac{2\phi}{\nu z} -  \frac{(d+z)}{z} =0.12.
\label{gammaSM}
\end{equation}	
In a small magnetic field H=0.03 T, $\chi(T)$ exhibits a powerlaw temperature dependence as shown in Fig. \ref{scalingb}b, over a small temperature range 1.8 K $<$ T $<$ 4.5 K, and yields the observed value $\gamma = 0.25$, close to the expected value of 0.12.

\section{Explicit Scaling of specific heat from Maxwell relation }

From the Maxwell relation  $\frac{\partial S}{\partial H} =  \frac{\partial M}{\partial T}$, we integrate with respect to field to give:
\begin{equation}
   \int_{0}^{H}\frac{\partial S}{\partial H}\, dH =  \int_{0}^{H}\frac{\partial M}{\partial T}\,dH. 	
\label{Integral}
\end{equation}	
Taking a temperature derivative on both sides, this simplifies to:
\begin{equation}
   \frac{\Delta C_{4f}}{T} = \frac{\partial^2}{\partial T^2} \int_{0}^{H}M(H,T) \,dH. 	
\label{Cp}
\end{equation}	
Using the explicit form of the scaling function for the magnetization
\begin{equation}
M =   H^{\frac{2-\alpha -\phi}{\phi}} F_1(x) = c(x^2+a^2)^{-\gamma/2}, \nonumber
\label{Magnetization}
\end{equation}	
we obtain the scaling form of the specific heat coefficient:
\begin{equation}
   \frac{\Delta C_{4f}}{T} = \frac{\partial^2}{\partial T^2} \int_{0}^{H}  H^{\frac{2-\alpha -\phi}{\phi}} F_1(x) \,dH = \int_{0}^{H} \frac{F_1^{\prime\prime}(x)}{H} \,dH,  	
\label{Cp_scaling}
\end{equation}	
\begin{equation} 
\mathrm{where} \,\,\, F_1^{\prime\prime} = \frac{\partial^2 F_1(x) }{\partial T^2}
= -c \gamma (x^2+a^2)^{-\gamma/2 -1}\Bigg[1 - (2+ \gamma)\frac{x^2}{x^2+ a^2}\Bigg]. \nonumber
\label{MagnSimple2}
\end{equation}	
This expression is used to obtain the scaling curve (red line) for the specific heat coefficient displayed in Fig. 2c.

\section{Phase Diagram near an AFMQCP}
\begin{figure*}[t]
\centerline{\includegraphics[width=6.5in]{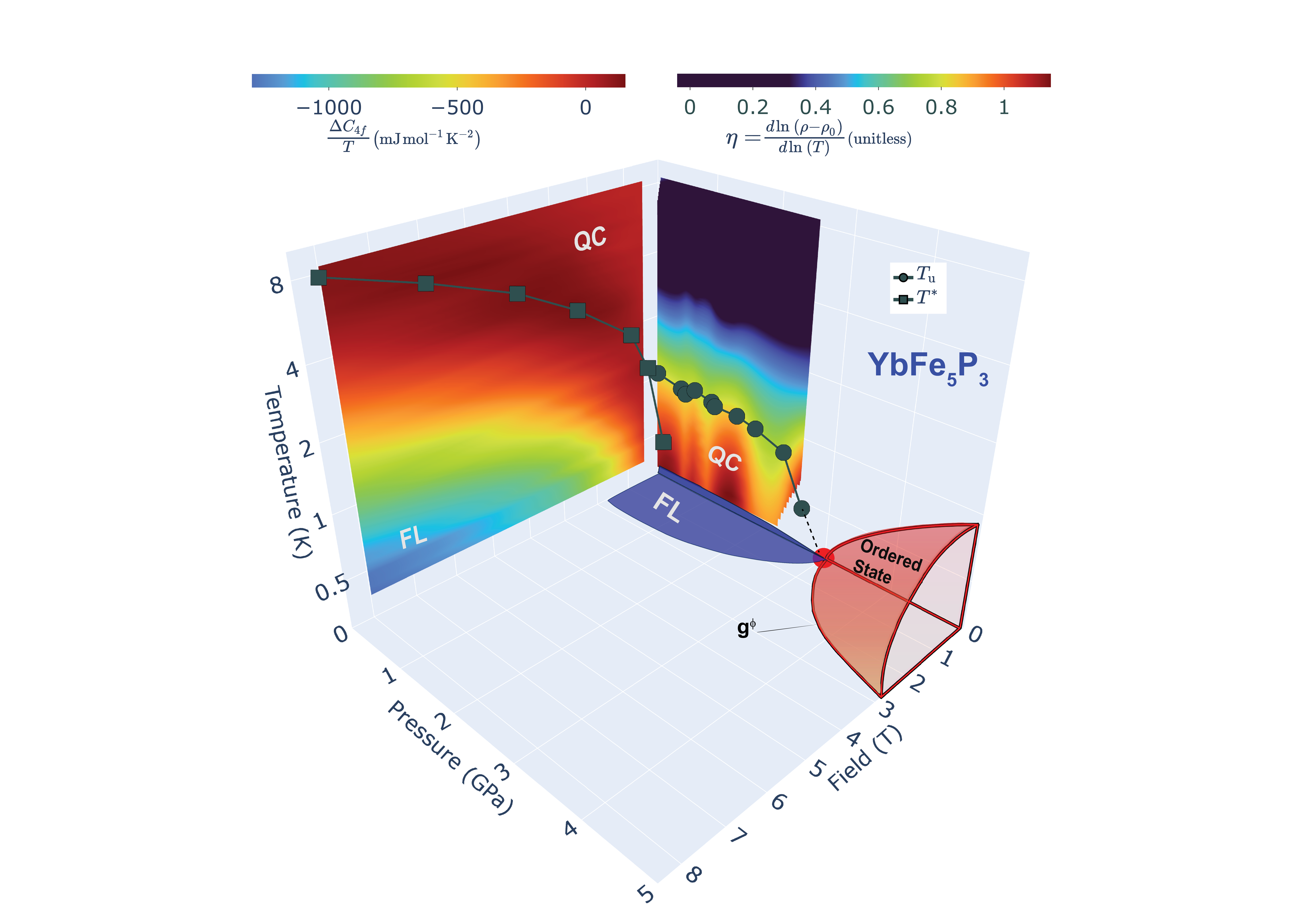}}
\caption{Temperature-Pressure-Magnetic Field ($T-P-H$) phase diagram of \ybfep.  The $T-H$ contour plot of $\Delta C_{4f}/T$ T shows the crossover from quantum critical to a Fermi liquid state.  The grey squares represent $T^\star (H)$, taken from the maximum in $\Delta C_{4f}/T$  at various magnetic fields (see Fig. 2d).  The $T-P$ contour plot is the powerlaw exponent $\eta$ of the electrical resistivity (see text for details).  The grey circles represent $T_U$, which is the upper bound of the T-linear fit range.  The red circle at $P_c$=3 GPa is the quantum critical point.  Magnetic order is expected beyond $P_c$ (red region), for which the $P-H$ phase boundary is associated with the exponent $\phi$.  A similar phase diagram is expected for alloys of \ybfep{} with Co and Ru, where the pressure axis is replaced by the solute concentration $y$.}
\label{phaseSM}
\end{figure*}
In Fig. \ref{phaseSM}, we plot the phase diagram for \ybfep{} as a function of pressure, temperature, and magnetic field. The color plot in the $H-T$ plane represents the magnitude of the specific heat $\Delta C_{4f}/T$. The grey squares are taken from Fig.  2d, and represent the crossover $T^\star(H)$ from the low-dimensional quantum critical (QC) state to the 3D Fermi liquid state (blue), which we believe to be the ultimate ground state in the disordered phases. The color data in the $P-T$ plane represents the power law exponent of the resistivity $\eta=dln(\rho -\rho_0)/dln(T)$,\cite{Asaba2021} Down to the lowest temperatures measured (~0.25 K), the resistivity is nearly linear ($\eta$ =1), a reflection of the quantum critical behavior. The black circles represent $T_U$, which is the upper bound of the T-linear fit range. Extrapolating $T_U\rightarrow 0$, a QCP is expected at a critical pressure $P_c \sim $3 GPa, beyond which magnetic order is expected (red region). A Fermi liquid state may exist below 0.25 K at low pressure, consistent with the FL state below $\sim$0.5 K at $P=0$ (Fig. 1b), and is eventually suppressed at $P_c$.  Replacing $P$ by alloy concentration $y$, such magnetic order is observed in the alloy Yb(Fe$_{1-y}$Co$_y$)$_5$P$_3$, where the QCP occurs at $y_c$=0.06; the transition is to a $\mathbf{k}$=(0,0,0) AFM state.\cite{Avers2024}

\section{Extended Mean-Field versus Gaussian behavior near a QCP }
We write the free energy of our system as a sum of $\mathbf{two}$ contributions: a mean-field part and a Gaussian one. The Gaussian part includes thermal and quantum fluctuations in the region of the phase diagram without long-range order. In mean field there are no fluctuations in this disordered (paramagnetic) phase, in the absence of a magnetic field.
\subsection{The extended mean-field approach}
	
\subsubsection{Ferromagnet (d = 1, 2, 3 and z = 3)} 

\begin{figure*}[ht]
\centerline{\includegraphics[width=4in,angle=270]{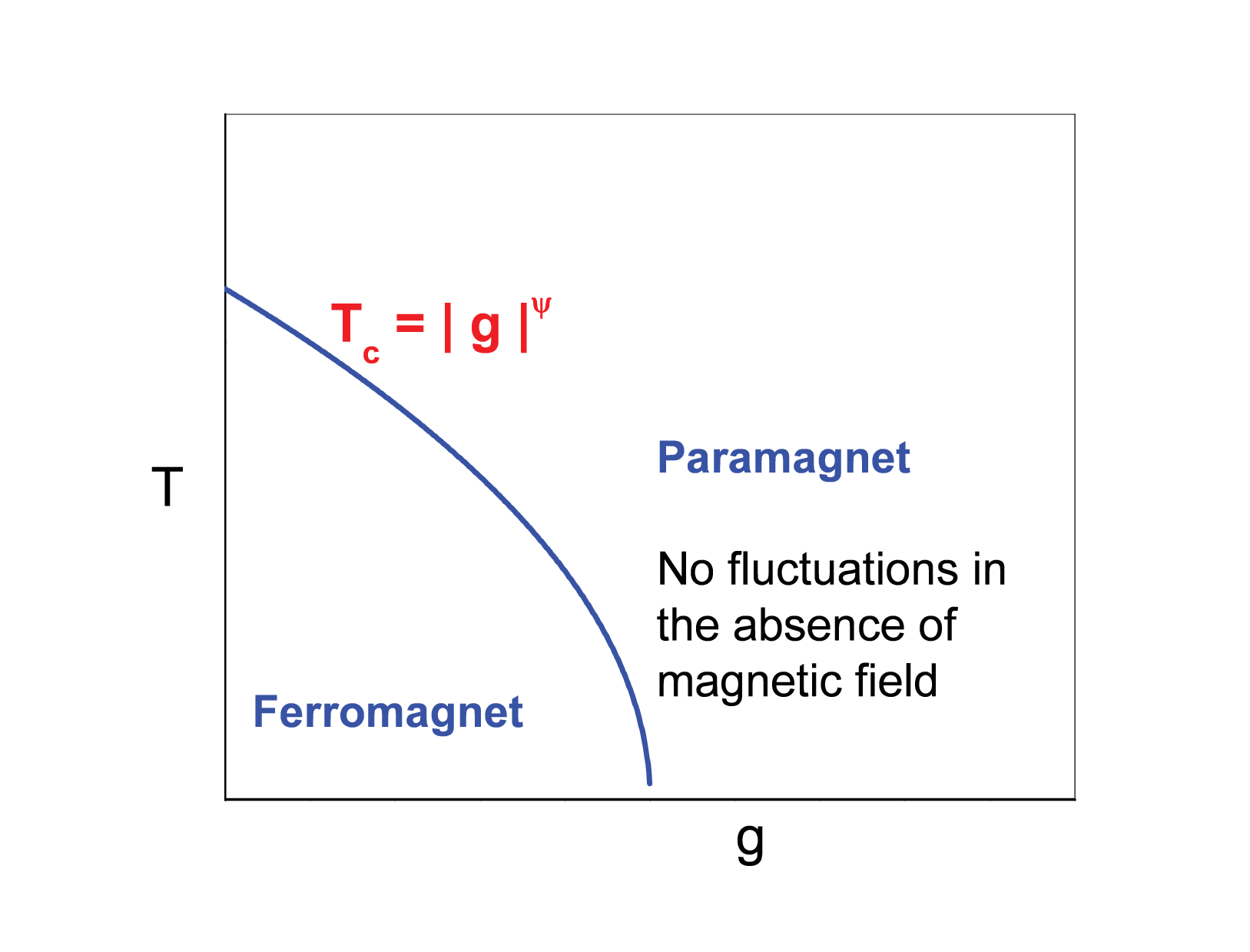}}
\caption{Phase diagram of the ferromagnet in the extended mean field approximation for $H=0$. In the absence of a magnetic field, the specific heat is zero in the paramagnetic phase because there are no fluctuations in the system.}
\label{Fig_FM_EMFA}
\end{figure*}

The scaling form of the free energy of a ferromagnet close to a quantum critical point in the extended mean-field approximation (EMFA) for $d+z \geq 4$ in the presence of a uniform magnetic field is written as, 
\begin{equation}
f = \left( \frac{|g|^{2-\alpha}}{u} \right) F_M\left[ \frac{H}{|g|^{\beta + \gamma}}\right],
\label{FE_FM}
\end{equation}
with
\begin{equation}
g=g_0- uT^{1/\psi}, 
\label{g0}
\end{equation}
and 
\begin{equation}
\psi =  z/(d+z-2)
\label{psi}
\end{equation}
where $g_0 = 0$ is the value of the driving parameter at the QCP.  All exponents in Eq.  \ref{FE_FM} are the mean field ones, i.e.,  $\alpha = 0, \beta =1/2,  \gamma =1$, and $F_M(H=0)=1$.   Eq. \ref{FE_FM} considers the existence of a critical line at finite temperatures and zero external magnetic field, described by a shift exponent $\psi$, that we take as that obtained by the renormalization group procedure of Millis,\cite{Millis93} In this EMFA,  the effects of the dynamics and dimensionality are implicitly considered in the shift exponent $\psi$. This exponent determines the shape of the critical line at finite temperatures close to the QCP, i.e., $T_C =(|(g_0  - g)/u|)^\psi$,   and arises due to the dangerous irrelevant character of the quartic interaction $u$ for $d+z \geq 4$, as obtained by Millis.\cite{Millis93}  Here, we consider the effect of $u$ also at $T=0$. The EMFA does not include fluctuations in the disordered, paramagnetic phase in the absence of an external magnetic field, where the order parameter vanishes, as shown in Fig. \ref{Fig_FM_EMFA}. Consequently, the specific heat in this approximation is zero in the disordered phase for $H=0$. The scaling form of the free energy, Eq. \ref{FE_FM} is consistent with the truncated Landau free energy expansion,\cite{Wu2014} 
\begin{equation}
\mathcal{F}=\chi_0^{-1} + M^2 + uM^4- HM, 
\label{chi_FM_FE}
\end{equation}
with $\chi_0^{-1}=g=g_0- uT^{1/\psi}$.

In EMFA, the magnetization is given by, 
\begin{equation}
M= |g|^{\beta} F_1\left[\frac{H}{|g|^{\beta+\gamma}}\right],
\end{equation}
At criticality ($g_0 =0$), using Eqs. \ref{FE_FM} and \ref{g0}, we get
\begin{equation}
M=H^{1/\delta} F_2 \left[ \frac{T}{|H|^{2\psi/3}} \right]
\end{equation}
where $ \delta =1+\gamma/\beta=3$ and $\alpha + 2\beta + \gamma =2$.
The uniform susceptibility is
\begin{equation}
    \chi = |g|^{-\gamma} F_3\left[\frac{H}{|g|^{\beta+\gamma}}\right],
\label{chi_FM}
\end{equation}	
At criticality ($g_0=0$),
\begin{equation}
    \chi =  |T|^{-\gamma/\psi} F_3\left[ \frac{T}{|H|^{2\psi/3}} \right]
\label{chi_FM_QCP}
\end{equation}	
where we used $\alpha + 2\beta + \gamma =2$, and the mean field values of the critical exponents.
We then have (for $z=3$):
\begin{eqnarray}
 \chi^{\mathrm{1D}} =  |T|^{-2/3} F_{\mathrm{1D}}\left[ \frac{T}{H} \right]  \\
\chi^{\mathrm{2D}} =  |T|^{-1} F_{\mathrm{2D}}\left[ \frac{T}{H^{2/3}} \right]  \\
\chi^{\mathrm{3D}} =  |T|^{-4/3} F_{\mathrm{3D}}\left[ \frac{T}{\sqrt{H}} \right].  \\
\end{eqnarray}
For the variation of the uniform susceptibility with temperature, we get at the QCP
\begin{equation}
    \frac{d\chi}{dT} = H^{-(2/3)(1+\psi)} F_4\left[ \frac{T}{|H|^{2\psi/3}} \right]
\end{equation}	
The scaling variable in the scaling functions for a ferromagnet with $z=3$ is given by
\begin{equation}
    \frac{H}{|T|^{(\beta + \gamma)/\psi}} = \frac{H}{|T|^{3/2\psi}} = \left( \frac{T}{H^{2/3}} \right)^{-3/2\psi},
\end{equation}	
or
\begin{equation}
\begin{cases}
 \frac{T}{H},& \mathrm{d=1} \\
\frac{T}{H^{2/3}},& \mathrm{d=2}  \\
 \frac{T}{\sqrt{H}},& \mathrm{d=3} 
\end{cases}
\end{equation}
The breakdown of $H/T$ scaling in two- and three-dimensions in a ferromagnet arises because the uniform magnetic field couples to the order parameter. 

Although the specific heat in the EMFA is zero in the disordered phase, there are fluctuations of the order parameter in the presence of a magnetic field and a contribution to the specific heat appears even in mean field. Following Ref. \onlinecite{Ma1985}, and using the fact that the free energy of a ferromagnet for a fixed finite field and temperature is a smooth function of $g$ (the QCP is at $H=T=0$), we write this free energy (Eq. \ref{FE_FM}) as,
\begin{equation}
\mathcal{F}=F_0(H) + F_1(H)g + ...,  
\label{FESM}
\end{equation}
Consistency with the scaling yields
\begin{eqnarray}
F_0(H) =H^{4/\delta}\\ 
F_1(H) =H^{2/\delta}
\end{eqnarray}
with the mean field exponent $\delta=3$. The next term in this expansion is field independent and corresponds to the mean-field energy in zero field. The specific heat obtained from the free energy expansion above is given by,
\begin{equation}
C_H/T=F_1(H)T^{\frac{1}{\psi}-2}.  
\end{equation}
For the ferromagnet with $z=3$, in $d=3$, we get, $C_H/T=(H)^{2/3} T^{-2/3}=(H/T)^{2/3}$. If $\psi=1/2$, as in the usual mean field case, we get a linear temperature dependent specific heat with a field dependent coefficient.

\subsubsection{Antiferromagnet, (z = 2, $d+ z>$ 4)} 
In this case, the uniform magnetic field is not conjugate to the order parameter, and we need to introduce a new exponent $\phi$ in the free energy:
\begin{equation}
f = \left( |g|^2 \right) F_5\left[ \frac{H}{|g|^{\phi}}\right],
\label{FE_AFM}
\end{equation}
where $g=g_0- uT^{1/\psi}$, and the shift exponent $\psi=z/(d+z-2)$ is different from that of the ferromagnetic case (z=3). If the uniform magnetic field is a relevant perturbation in the RG sense, then we refer to $\phi$ as a crossover exponent. At quantum criticality ($g_0=0$), 
\begin{equation}
f=\left(uT^{1/\psi}\right)^2 F_6\left[ \frac{H}{\left(uT^{1/\psi}\right)^\phi}\right].
\label{FE_AFM2}
\end{equation}
The uniform susceptibility for $H\rightarrow0$, is given by
\begin{equation}
    \chi =  \left(u^2 T^{2/\psi}\right) \left((1/u) T^{-2\phi}\right) =  (1/u)T^{\frac{2(1-\phi)}{\psi}}.
\label{chi_AFM2}
\end{equation}	
Because the uniform susceptibility does not vanish or diverge at the N\'{e}el transition of the antiferromagnet along $g=0$, we have $\phi=1$ in this mean-field approach. This yields for the magnetization and magnetic susceptibility at criticality,
\begin{eqnarray}
    M = H F_7\left[ \frac{H}{T^\psi}\right]\\
    \chi =  F_8\left[ \frac{H}{T^\psi}\right]
\end{eqnarray}
Notice that the relevant scaling variable is,
\begin{equation}
     H(T^{-\phi/\psi})=\left(T/H^{\psi/\phi} \right)^{-\psi/\phi}.
\end{equation}
that yields
\begin{equation}
\begin{cases}
  T/H,&   d=2\\
T/H^{2/3},&  d=3
\end{cases}
\end{equation}
where we used $\phi=1$, and $\psi = z/(d+z-2)$, with $z=2$ for an antiferromagnet.

\subsection{Gaussian Approximation}
\subsubsection{Ferromagnet, d +z $>$ 4}
The extended mean field approach discussed above does not consider fluctuations in the disordered phase in the absence of a magnetic field. The reason is that the order parameter is zero in this phase and consequently the mean field Landau free energy vanishes. The next approximation that includes fluctuations is the Gaussian free energy with quartic corrections. Notice that this is the correct approach for d+z $>$4, as the quartic interaction $u$ becomes dangerously irrelevant. In the case of a ferromagnet, the free energy in this Gaussian case is
\begin{equation}
f_G=|g^{\nu (d+z)}| F_G\left[ \frac{T}{|g|^{\nu z}}, \frac{H}{|g|^{\beta + \gamma}}\right] ,        
\end{equation}
where $g=g_0- uT^{1/\psi}$. The other critical exponents take mean field values.  Notice that at criticality, with $g_0 = 0$, we can write
\begin{equation}
\frac{T}{|g|^{\nu z}} = T \xi^z         
\end{equation}
where the correlation length $\xi = [(uT^{1/\psi}]^{-\nu}$, or $\xi^{-2}=uT^{1/\psi}$ (Ref. \onlinecite{Millis93}, Eq. 3.11)).
In the Gaussian approximation, for a nearly ferromagnetic metal, a Fermi liquid regime appears below the coherence line, $T_{coh}=|g|^{\nu z}$.\cite{Continentino2001} The coefficient of the linear term in the specific heat at zero field is enhanced as
\begin{equation}
C/T = |g_0|^{\nu(d- z)},          
\end{equation}
and at criticality the dominant Gaussian contribution yields,
\begin{equation}
C/T = T^{(d- z)/z}.          
\end{equation}
The field-dependent contributions in the Gaussian approximation are less singular than those obtained within the extended mean-field. To see that, consider the limit $T \rightarrow 0$, and at quantum criticality ($g_0$ = 0). In this case, for $d+z> 4$, all physical quantities obtained from field derivatives of the free energy are more singular in the extended mean field case than in the Gaussian case. For example, the zero-field uniform susceptibility of the 3D ferromagnet obtained from EMFA (Eq. \ref{FE_FM}), at quantum criticality is given by, 
\begin{equation}
\chi_{MF} = T^{-\gamma/\psi},          
\end{equation}
with $\gamma=1$ and $\psi=3/4$. In the Gaussian case, we get
\begin{equation}
\chi_{G} = T^{\nu(d- z) -2(\beta + \gamma)/\psi},          
\end{equation}
where  $\nu(d- z) -2(\beta + \gamma)/\psi = 0 $, such that $\chi_G$ does not diverge. The reason the mean field result is more singular is the breakdown of hyperscaling [$2-\alpha = \nu (d+z)$] for $d+z > 4$. In the mean field approximation, the exponent $\alpha= 0$ for any $d+z\geq 4$, while in the Gaussian case $\alpha = 2-\nu (d+z)<0$. 

\section{Other Possibilities for the QCP in YbFe$_5$P$_3$}
\subsection{Local Quantum Critical Point}
The Gaussian free energy for a nearly antiferromagnetic system can be written as (Ref. \onlinecite{Moriya95}):
\begin{equation}
f=\frac{-3}{\pi} \sum_q \int_0^\infty \frac{d\omega}{e^{\omega/T} - 1} \mathrm{tan}^{-1}\left[\frac{\omega + H}{|g|+(J_{Q+q}-J_q)}\right],         
\end{equation}
where  $J_q$ is the Fourier transform of the interactions between moments. Expanding $J_{Q+q}-J_q= Aq^2+...$, we rewrite $f$ as,
\begin{equation}
f=\frac{-3}{\pi} \sum_q \int_0^\infty \frac{d\omega}{e^{\omega/T} - 1} \mathrm{tan}^{-1}\left[\frac{\lambda (T/T_{coh}) + (H/T_{coh})}{(1 + (q \xi)^2)}\right],         
\end{equation}
with $T_{coh}=|g|^{\nu z}$, $\xi =\sqrt{A/|g|}$, and $\nu =1/2$ and $z=2$, for an AFMQCP. It is easy to see that below the coherence temperature ($T \ll T_{coh}$), the free energy is quadratic in temperature signaling that the system is in the Fermi liquid regime.  The local quantum critical regime satisfies the condition,\cite{Continentino2001}
$q_c^2 \xi^2 \ll 1$,  or $|g|\gg A q_c^2$,
where $q_c$ is a cut-off in momentum space.    Notice that when the stiffness $A\rightarrow 0$, the local regime may extend arbitrarily close to the AFQCP. In this local regime, we have a single parameter scaling, namely, the coherence temperature $T_{coh}$. This is different from the Kondo impurity problem, as the characteristic temperature here is a collective property that depends on the distance to the QCP.

The scaling form of the free energy in the local regime is given by
\begin{equation}
f_G=|g^{\nu (d+z)}| F\left[ \frac{T}{T_{coh}}, \frac{H}{T_{coh}}\right],        
\end{equation}
with d=0. Notice that this obeys $T/H$ scaling. The magnetization, $M=F[T/H]$ and the uniform susceptibility increases with decreasing temperature as $\chi = (1/T)F[T/H]$.

\subsection{Lifschitz Quantum Criticality}
This case has been discussed in Refs. \onlinecite{Ramazashvili1999,Continentino2004}. When the stiffness $A=0$, we must go to the next term in the expansion of $J_{Q+q}-J_q$. The Gaussian free energy is given by,
\begin{equation}
f=\frac{-3}{\pi} \sum_q \int_0^\infty \frac{d\lambda}{e^{\lambda} - 1} \mathrm{tan}^{-1}\left[\frac{\lambda (T/T_{coh}) + (H/T_{coh})}{(1 + (B q \xi_L)^4)}\right],         
\end{equation}
where $T_{coh}=|g|^{\nu_L z_L}$  and $\xi_L=|g|^{-\nu_L}$ with $\nu_L=1/4$ and  $z_L$=4.  This free energy obeys the scaling form,
\begin{equation}
f=|g|^{\nu_L (d+z_L)} F\left[ \frac{T}{T_{coh}}, \frac{H}{T_{coh}}\right].        
\end{equation}
In $d=2$, we get
\begin{equation}
f=|g|^{3/2} F\left[ \frac{T}{T_{coh}}, \frac{H}{T_{coh}}\right].  
\end{equation}
or
\begin{equation}
f=|H|^{3/2} F\left[ \frac{T}{H}\right].  
\end{equation}
Again, we have $T/H$ scaling. Notice that this is exactly the scaling form of the free energy in $\beta$-YbAlB$_4$ (Ref. \onlinecite{Matsumoto2011}, Eq. (2)), from which results all the scaling properties of the field dependent data in this compound.

\section{Scaling Results for CeRu$_6$Sn$_4$}

We use the scaling results of CeRu$_6$Sn$_4$ presented in Fig. 2D for $H||c$.\cite{Fuhrman2021}  In the hyperscaling regime ($d+z <4$), the application of Eq. 9 in the main text leads to the conclusion:    
\begin{align}
        & \nu z/\phi=0.43,\nonumber\\ 
   & 2- \frac{\nu d}{\phi}=0.78. 
    \label{exponents_Ce}
\end{align}
 This implies that $z/d=0.35$ and $\nu/\phi =0.43/z$.  This gives the scaling parameter predictions list in Table \ref{Table1}.
 We note that only the ratio $\nu/\phi$ is determined; without further information, $\nu$ and $\phi$ cannot be separately determined. We note that the values of $z$ for d=1 and d=2 are quite unusual, but for d=3, $z$ is nearly equal to unity. Assuming $z=1$ gives $z+d=4$, where the mean field equations should also apply. For 
H = 0, we have $\theta = \psi = z/(d+z-2)$ and $\alpha$ = 0. Using Eq. 5 in the main text, we have $\psi/\phi$= 0.43, and 	$2 +  [(\psi-2)/\phi] = 0.78$. Solving these equations gives
$\psi  = 0.52$ and $\phi$ = 1.21. With $\psi \approx  1/2$, then for d=3, where $\psi = z/(z+1)$, we have $z=1$.  We note that $\psi/\phi$ is essentially identical to the exponent $\nu z / \phi$ for the hyperscaling case with d = 3. Similarly, $z=1$ in both cases; we have discussed the possibility that $z=1$ in the main text. Also, the value of $\phi$ is very similar to those reported for random antiferromagnets.\cite{Ferreira1991,Fishman1979}  
\begin{table}[h]
     \centering
     \begin{tabular}{c|c|c}
          & z & $\nu /\phi$\\\hline
 \textbf{d=1}          & 0.35 & 1.43 \\\hline
  \textbf{d=2}          & 0.71 &  0.61 \\\hline
    \textbf{d=3}          & 1.06 &  0.41 \\
     \end{tabular}
     \caption{Scaling parameters ($z$ and $\nu /\phi$) of CeRu$_6$Sn$_4$, taken from results presented in Fig. 2D of Ref. \onlinecite{Fuhrman2021}, for the d=1,2, and 3 cases. }
     \label{Table1}
 \end{table}

Finally, we point out that in the mean field regime where $\nu= 1/2$, given  $z = 1$, we also have $\nu z = 1/2$. According to Millis,\cite{Millis93} the thermal evolution of the system in the quantum critical regime is governed by the exponent $\nu_T = \nu/\psi$. Because $\psi = 1/2$, we obtain $\nu_T = 1$. The exponent $\nu_T z$, which is unity here, governs the evolution of the energy scale $\omega \sim T^{\nu_T z}$.  Hence, this analysis is consistent with the observed $\omega/T$ scaling in the neutron scattering experiment on CeRu$_6$Sn$_4$.\cite{Fuhrman2021} Thus, these scaling results suggest that a QCP with 3D fluctuations with $z=1$ is possible.

\end{document}